\documentclass[superscriptaddress,aps,preprintnumbers,showpacs,prd,nofootinbib,reprint]{revtex4-1}
\usepackage{xcolor}
\usepackage{amssymb,amsthm,amsmath,bm,slashed,graphicx,soul}
\usepackage[caption=false]{subfig}
\usepackage{multirow}
\usepackage{tabularx}
\usepackage[colorlinks,allcolors=blue]{hyperref} 
\usepackage[normalem]{ulem}
\begin{document}


\renewcommand{\figurename}{Fig.}
\renewcommand{\tablename}{Table.}
\newcommand{\Slash}[1]{{\ooalign{\hfil#1\hfil\crcr\raise.167ex\hbox{/}}}}
\newcommand{\bra}[1]{ \langle {#1} | }
\newcommand{\ket}[1]{ | {#1} \rangle }
\newcommand{\bef}{\begin{figure}}  \newcommand{\eef}{\end{figure}}
\newcommand{\bec}{\begin{center}}  \newcommand{\eec}{\end{center}}
\newcommand{\laq}[1]{\label{eq:#1}}  
\newcommand{\dd}[1]{{d \o d{#1}}}
\newcommand{\Eq}[1]{Eq.~(\ref{eq:#1})}
\newcommand{\Eqs}[1]{Eqs.~(\ref{eq:#1})}
\newcommand{\eq}[1]{(\ref{eq:#1})}
\newcommand{\Sec}[1]{Sec.\ref{chap:#1}}
\newcommand{\ab}[1]{\left|{#1}\right|}
\newcommand{\vev}[1]{ \left\langle {#1} \right\rangle }
\newcommand{\bs}[1]{ {\boldsymbol {#1}} }
\newcommand{\lac}[1]{\label{chap:#1}}
\newcommand{\SU}[1]{{\rm SU{#1} } }
\newcommand{\SO}[1]{{\rm SO{#1}} }
\def\({\left(}
\def\){\right)}
\def\dt{{d \o dt}}
\def\diag{\mathop{\rm diag}\nolimits}
\def\Spin{\mathop{\rm Spin}}
\def\O{\mathcal{O}}
\def\U{\mathop{\rm U}}
\def\Sp{\mathop{\rm Sp}}
\def\SL{\mathop{\rm SL}}
\def\tr{\mathop{\rm tr}}
\def\ebq{\end{equation} \begin{equation}}
\newcommand{\OR}{~{\rm or}~}
\newcommand{\AND}{~{\rm and}~}
\newcommand{\EV}{ {\rm \, eV} }
\newcommand{\KEV}{ {\rm \, keV} }
\newcommand{\MEV}{ {\rm \, MeV} }
\newcommand{\GEV}{ {\rm \, GeV} }
\newcommand{\TEV}{ {\rm \, TeV} }
\def\o{\over}
\def\a{\alpha}
\def\b{\beta}
\def\c{\varepsilon}
\def\d{\delta}
\def\e{\epsilon}
\def\f{\phi}
\def\g{\gamma}
\def\h{\theta}
\def\k{\kappa}
\def\l{\lambda}
\def\m{\mu}
\def\n{\nu}
\def\p{\psi}
\def\q{\partial}
\def\r{\rho}
\def\s{\sigma}
\def\t{\tau}
\def\u{\upsilon}
\def\w{\omega}
\def\x{\xi}
\def\y{\eta}
\def\z{\zeta}
\def\D{\Delta}
\def\G{\Gamma}
\def\H{\Theta}
\def\L{\Lambda}
\def\F{\Phi}
\def\P{\Psi}
\def\S{\Sigma}
\def\me{\mathrm e}
\def\ol{\overline}
\def\tl{\tilde}
\def\*{\dagger}

\preprint{TU-1221}

\title{ 
Bubble Misalignment Mechanism for Axions
}

\author{
Junseok Lee
}
\affiliation{Department of Physics, Tohoku University, 
Sendai, Miyagi 980-8578, Japan} 

\author{
Kai Murai
}
\affiliation{Department of Physics, Tohoku University, 
Sendai, Miyagi 980-8578, Japan} 

\author{
Fuminobu Takahashi
}
\affiliation{Department of Physics, Tohoku University, 
Sendai, Miyagi 980-8578, Japan} 

\author{
Wen Yin
}
\affiliation{Department of Physics, Tohoku University, 
Sendai, Miyagi 980-8578, Japan}

\begin{abstract}
We study the dynamics of axions at first-order phase transitions in non-Abelian gauge theories.
When the duration of the phase transition is short compared to the timescale of the axion oscillations, the axion dynamics is similar to the trapped misalignment mechanism. On the other hand, if this is not the case,  the axions are initially expelled from the inside of the bubbles, generating axion waves on the outside. Analogous to the Fermi acceleration, these axions gain energy by repeatedly scattering off the bubble walls. Once they acquire enough energy, they can enter the bubbles. The novel ``bubble misalignment mechanism'' can significantly enhance the axion abundance, compared to models where the axion mass is either constant or varies continuously as a function of temperature. The increase in axion abundance depends on the axion mass, the duration of the phase transition, and the bubble wall velocity.
This mechanism results in a spatially inhomogeneous distribution of axions, which could lead to the formation of axion miniclusters.
It has potential implications for 
the formation of oscillons/I-balls, axion warm dark matter, cosmic birefringence,
and the production of dark photons.
\end{abstract}

\maketitle
\flushbottom

\vspace{1cm}

\section{Introduction}
\label{sec: intro}
The axion is a pseudo Nambu-Goldstone boson that acquires mass due to the breaking of its shift symmetry. For instance, in the Peccei-Quinn mechanism, a solution to the strong CP problem, the QCD axion gains mass through the non-perturbative effects of QCD~\cite{Peccei:1977hh,Peccei:1977ur,Weinberg:1977ma,Wilczek:1977pj}. In particular, its mass is significantly suppressed at high temperatures but increases as the temperature approaches the QCD scale, eventually asymptoting to a constant value. More general axions are considered to similarly acquire mass through non-perturbative effects of non-Abelian gauge fields in hidden sectors. In this paper, we discuss the dynamics of axions in scenarios where their mass changes discontinuously due to a first-order phase transition (FOPT).

Dark matter is direct evidence of physics beyond the Standard Model, and its true nature remains elusive. The axion is one of the promising candidates for dark matter. The misalignment mechanism~\cite{Preskill:1982cy,Abbott:1982af,Dine:1982ah} is a known mechanism for the generation of axion dark matter; the axion begins oscillating around the potential minimum when its mass approximately equals the Hubble parameter, and its oscillation energy can account for dark matter. Various extensions have been considered for the misalignment mechanism. For example, a trapped misalignment mechanism where the oscillation onset is delayed due to the trapping potentials~\cite{Higaki:2016yqk,DiLuzio:2021gos,Jeong:2022kdr}, the initial misalignment angle around a special position realized by axion mixings~\cite{Daido:2017wwb,Takahashi:2019pqf,Nakagawa:2020eeg,Narita:2023naj},
resonances between multiple axions~\cite{Kitajima:2014xla,Daido:2015bva,Daido:2015cba,Ho:2018qur,Murai:2023xjn,Nakagawa:2022wwm,Cyncynates:2023esj},
and the kinetic misalignment mechanism based on non-trivial radial dynamics and explicit breaking of U(1)$_{\rm PQ}$ symmetry~\cite{Co:2019jts}.

Recently, Nakagawa, Yamada, and two of the authors (FT and WY) proposed a misalignment mechanism for scenarios where the axion mass arises from a coupling to non-Abelian gauge fields that undergo a FOPT from a deconfined to a confined phase~\cite{Nakagawa:2022wwm}. For instance, it is known that in the pure SU($N$) Yang-Milles theory, the transition becomes first order for $N\geq 3$~\cite{Lucini:2003zr,Lucini:2005vg}, making this a plausible scenario. Interestingly, the previous study of the misalignment mechanism based on FOPTs showed a significant increase in axion abundance compared to scenarios where the axion mass is constant or a continuous function of temperature~\cite{Nakagawa:2022wwm}. This mechanism results in a delayed onset of oscillations. However, unlike trapped misalignment mechanisms, bubble formation and coalescence in the FOPT create spatial inhomogeneity, an effect that has not been considered in previous literature. This paper aims to explore the dynamics of axions considering bubble formation and coalescence, an essential element in FOPTs.

We find that axion dynamics are greatly affected by the difference in axion mass between the inside and outside of the bubbles, as well as by the velocity of the bubble wall.
In particular, when the mass difference is relatively large and the time scale of the axion oscillations inside the bubbles is shorter than the duration of the FOPT, axions are initially expelled from the inside of the bubble.
The scattered axion waves propagate outside the bubbles and then scatter off another bubble.
Consequently, they gain energy by repeatedly scattering with bubbles, similar to the Fermi acceleration mechanism~\cite{Blandford:1978ky,Bell:1978clk,Drury:1983zz,Blandford:1987pw}.
As they become more energetic, axions begin to enter the bubbles, keeping the number of axions. On the other hand, if the duration of FOPT is shorter than the time scale of the axion oscillations, the oscillation amplitude of the axion remains unchanged during the FOPT. In this case, the axion abundance is enhanced by the ratio of the mass squared, reproducing the results of Ref.~\cite{Nakagawa:2022wwm}. We call the mechanism of axion production by bubble nucleation in FOPT the ``bubble misalignment mechanism,'' and in this paper, we clarify its dynamics and discuss its cosmological implications.

{Let us comment on some early studies of dark matter production scenarios related to bubble wall dynamics during FOPT. In Refs.~\cite{Baker:2019ndr,Chway:2019kft}, the authors consider the dark matter plasma in thermal equilibrium in the symmetric phase, while in the broken phase, the dark matter becomes heavy. As the bubble expands, most dark matter particles are reflected at the walls, and only Boltzmann-suppressed high-energy particles can enter the broken phase, suppressing the heavy dark matter number density. 
In Refs.~\cite{Azatov:2021ifm, Baldes:2022oev, Azatov:2022tii} it was pointed out that the boosted heavy dark matter can be produced via the reaction of the thermal plasma in the symmetric phase with the Higgs field for the bubble wall, a process which can produce the heavy dark matter more efficiently than the collisions of the bubble walls
\cite{Falkowski:2012fb}. In these studies, as usually assumed in the analysis, the wall width in the wall rest frame is taken to be larger or comparable to the plasma de Broglie wavelength see, e.g., \cite{Bodeker:2009qy,Bodeker:2017cim,Hoche:2020ysm,Azatov:2020ufh}. Namely, the plasma or dark matter is treated as a particle. 
In the context of the friction to the bubble wall, the boundary condition of low momentum modes of the gauge field was discussed in Refs.~\cite{GarciaGarcia:2022yqb,Azatov:2023xem}, but the gauge field is not a very weakly coupled field.

The main difference of this paper from the previous works is that we consider the bubble wall width in the wall rest frame to be much smaller than the de Broglie wavelength of the out-of-equilibrium and weakly coupled scalar field, i.e., the dark matter is a wave interacting with the wall. 
This leads to a significantly different analysis and phenomena than those discussed in previous works. In particular, phenomena such as Fermi acceleration, shock waves, and enhancement of dark matter abundance are not observed in the references. 
Our mechanism is mainly for light and wavy dark matter rather than heavy and particle dark matter.

The rest of this paper is organized as follows. In Sec.~\ref{sec: FOPT & axion potential} we briefly describe the nature of bubbles and axion potentials in FOPTs and clarify the setup of the problem we consider. In Sec.~\ref{sec: spatially uniform} we first discuss the axion abundance in the spatially uniform case, ignoring the bubbles, for comparison with the literature. In Sec.~\ref{sec: axion wave dynamics}, after classifying the axion dynamics, the Fermi acceleration process by the reflection and transmission of axion waves between the bubble walls is discussed in detail. In Sec.~\ref{sec: bubble misalignment}, based on the discussion in the previous section, we evaluate the axion abundance for each case. The last section is reserved for a discussion and conclusions.

\section{FOPT and axion potential}
\label{sec: FOPT & axion potential}
In FOPTs, bubbles of the true vacuum nucleate, expand, and collide with each other.
Here, we characterize the dynamics of bubbles by the bubble wall velocity $v$, the bubble nucleation temperature $T_{\mathrm{b}}$, and the inverse of the duration of the FOPT $\beta$.\footnote{In this paper, we ignore the time difference between when the bubble nucleation rate exceeds the Hubble parameter and when the bubbles percolate, and denote the temperature at these times collectively as $T_{\rm b}$.}
Then, the mean separation of bubbles is given by $R \simeq v/\beta$.
These parameters depend on models of FOPT. {See e.g. Ref.~\cite{Gouttenoire:2023roe}. In the following, we treat $v$, $T_\mathrm{b}$, and $\beta$ as free parameters to describe a wide range of FOPT models.}

The axion potential is typically given by
\begin{align}
    V(\phi)
    &=
    \chi(T) \left[ 
        1 - \cos \left( \frac{\phi}{f_\phi} \right)
    \right]
    \ ,
    \label{eq: potential}
\end{align}
where $f_\phi$ is the decay constant of $\phi$, and we have chosen the origin of $\phi$ so that $\phi = 0$ becomes the potential minimum.
In the false vacuum, the topological susceptibility, $\chi(T)$, depends on the temperature as 
\begin{align}
    \chi(T)
    \simeq
    \left\{
        \begin{array}{ll}
            \chi_0 
            & \quad
            (T < \Lambda)
            \\
            \chi_0 \displaystyle{\left( \frac{T}{\Lambda} \right)^{-p} }
            & \quad
            (T \geq \Lambda)
        \end{array}
    \right.
    \ .
    \label{eq: chi}
\end{align} 
On the other hand, in the true vacuum, $\chi(T)$ takes a constant value $\chi(T) = \chi_0$.
We show the temperature dependence of $\chi(T)$ in Fig.~\ref{fig: chi}.
Such behavior of the topological susceptibility has been observed in the numerical lattice simulations of the SU(3) Yang-Milles theory~\cite{Borsanyi:2022fub}.
\begin{figure}[!t]
    \begin{center}  
        \includegraphics[width=0.45\textwidth]{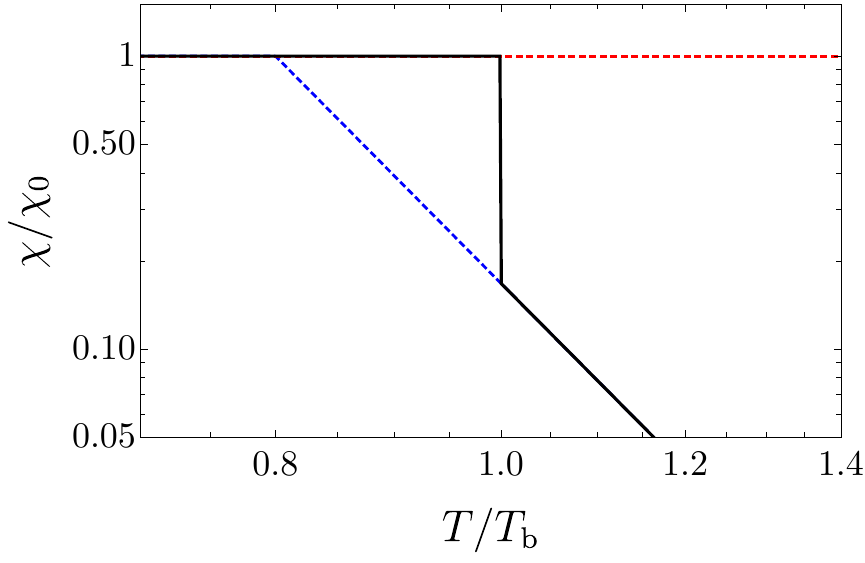}
        \end{center}
    \caption{%
        Temperature dependence of $\chi(T)$ for {$\Lambda/T_{{\mathrm{b}}} = 0.8$ and $p = 8$.}
        The blue and red dashed lines represent $\chi$ in the false and true vacua, respectively.
    }
    \label{fig: chi} 
\end{figure}
For later convenience, we define the axion mass parameters by
\begin{align}
    m_\phi(T) 
    \equiv 
    \frac{\sqrt{\chi(T)}}{f_\phi}
    \ , \quad 
    m_{{\mathrm{b}}} 
    \equiv 
    \frac{\sqrt{\chi(T_{\mathrm{b}})}}{f_\phi}
    \ , \quad 
    m_{0} 
    \equiv 
    \frac{\sqrt{\chi_0}}{f_\phi}
    \ ,
\end{align}
and treat them as free parameters, but with the assumption that $m_0 > m_{\rm b}$.

\section{Analysis of the spatially uniform case}
\label{sec: spatially uniform}

The discontinuous change of the axion mass at the FOPT affects the axion abundance in the later universe.
Here, we ignore the bubble dynamics and assume that the FOPT occurs uniformly in the universe.
Then, we can divide the axion dynamics into 
{three cases: $m_\mathrm{b} > 3H_\mathrm{b}$, $m_\mathrm{0} > 3H_\mathrm{b} > m_\mathrm{b}$, and $3H_\mathrm{b} > m_0$.
Here, $H_\mathrm{b} \equiv H(T_\mathrm{b})$ is the Hubble parameter at $T = T_\mathrm{b}$.
In the third case, the axion starts to oscillate after the FOPT, and the axion dynamics is the same as in the case where the axion has a constant mass $m_0$.}

In the first case of $m_\mathrm{b} > 3H_\mathrm{b}$, the axion starts to oscillate before the nucleation of bubbles.
The temperature of the onset of oscillations, $T_\mathrm{osc}$, is estimated by
\begin{align}
    m_\phi(T_\mathrm{osc}) 
    =
    3H(T_\mathrm{osc})
    \ ,
\end{align}
which leads to 
\begin{align}
    T_\mathrm{osc}
    &\simeq 
    \left( 
        \sqrt{\frac{10}{\pi^2 g_*(T_\mathrm{osc})}}
        m_0 M_\mathrm{Pl} \Lambda^{p/2} 
    \right)^{\frac{2}{p+4}}
    \ .
\end{align}
Here, we used the Friedmann equation during the radiation-dominated era,
\begin{align}
    3 M_\mathrm{Pl}^2 H^2 
    =
    \frac{\pi^2}{30}g_*(T) T^4
    \ ,
\end{align}
with $M_\mathrm{Pl} \simeq 2.4 \times 10^{18}$\,GeV being the reduced Planck mass and $g_*$ being the relativistic degrees of freedom for the energy density.
After then, the axion oscillates with a decreasing amplitude, $\bar{\phi}$.
Until the phase transition, $\chi(T)$ evolves much slower than the oscillations of $\phi$.
Then, the comoving number density of $\phi$ is conserved, and  $\bar{\phi}$ is approximately given by
\begin{align}
    \bar{\phi} 
    \simeq 
    \phi_0 
    \sqrt{\frac{m_\phi(T_\mathrm{osc})}{m_\phi(T)}}
    \left( \frac{a_\mathrm{osc}}{a} \right)^{3/2}
    \ ,
\end{align}
where $a$ is the scale factor, and the subscript `osc' implies that the variable is evaluated at $T = T_{\rm osc}$.
Ignoring the time evolution of the relativistic degree of freedom for the entropy density, $g_{*s}$, we obtain
\begin{align}
    \bar{\phi}(T)
    \simeq 
    \phi_0 
    \left( \frac{T}{T_\mathrm{osc}} \right)^{\frac{p+6}{4}}
    \ .
    \label{eq: axion amplitude for growing m}
\end{align}
Consequently, $\phi$ can be approximated by 
\begin{align}
    \phi
    \sim 
    \phi_0 
    \left( \frac{T}{T_\mathrm{osc}} \right)^{\frac{p+6}{4}}
    \sin \left[ \int \mathrm{d}t \, m_\phi(T) \right]
    \ .
\end{align}

At the FOPT, we can approximate $\phi$ and $\dot{\phi}$
by
\begin{align}
    \phi 
    &\simeq 
    \bar{\phi}_\mathrm{b} \sin \alpha_\mathrm{b}
    \ , \\
    \dot{\phi} 
    &\simeq 
    m_\mathrm{b} \bar{\phi}_\mathrm{b} \cos \alpha_\mathrm{b}
    \ ,
\end{align}
where $\bar{\phi}_\mathrm{b} \equiv \bar{\phi}(T_\mathrm{b})$, and $\alpha_\mathrm{b}$ 
{represents a phase of oscillations at the FOPT.}
Since the axion field oscillates with a mass of $m_0$ after the FOPT, the energy density of $\phi$ just after the FOPT is given by 
\begin{align}
    \rho_\phi(T_\mathrm{b})
    =
    \frac{\bar{\phi}_\mathrm{b}^2}{2}
    \left( 
        m_\mathrm{b}^2 \cos^2 \alpha_\mathrm{b}
        +
        m_0^2 \sin^2 \alpha_\mathrm{b}
    \right)
    \ .
\end{align}
{Note that this assumes that both $\phi$ and $\dot{\phi}$ are preserved in FOPT. The condition for this will be clarified in the next section.}
One can see that, depending on $\alpha_\mathrm{b}$, $\rho_\phi$ can be enhanced at the phase transition.
In particular, it is maximally enhanced by a factor of $m_0^2/m_\mathrm{b}^2$ when $\cos \alpha_\mathrm{b} = 0$, but it remains unchanged at the phase transition if $\sin \alpha_\mathrm{b} = 0$.

In the second case of $m_\mathrm{0} > {3 H_\mathrm{b}} 
> m_\mathrm{b}$, the axion remains constant until the phase transition and begins to oscillate immediately afterward. Therefore, the energy density just after the phase transition is 
\begin{align}
    \rho_\phi(T_\mathrm{b})
    =
    \frac{m_0^2 \phi_0^2}{2}
    \ .
\end{align}
{The sudden onset of oscillations has some similarities to the so-called trapped misalignment mechanism~\cite{Higaki:2016yqk,DiLuzio:2021gos,Jeong:2022kdr}.}

For comparison, we also consider the case of the second-order phase transition (or crossover) and evaluate the axion energy density for $T < \Lambda$.
Here, we assume that the topological susceptibility follows Eq.~\eqref{eq: chi}.
Then, the oscillation amplitude of the axion is given by Eq.~\eqref{eq: axion amplitude for growing m} for $T > \Lambda$
and by
\begin{align}
    \bar{\phi}(T)
    \simeq 
    \phi_0 
    \left( \frac{\Lambda}{T_\mathrm{osc}} \right)^{\frac{p+6}{4}}
    \left( \frac{T}{\Lambda} \right)^{3/2}
\end{align}
for $T < \Lambda$, where we ignored the temperature dependence of $g_{*s}$.
Thus, the axion energy density is given by 
\begin{align}
    \rho_\phi^{\mathrm{2nd}}(T)
    \simeq 
    \frac{m_0^2 \phi_0^2}{2}
    \left( \frac{\Lambda}{T_\mathrm{osc}} \right)^{\frac{p+6}{2}}
    \left( \frac{T}{\Lambda} \right)^3
    \ .
\end{align}
We obtain the ratio of the axion energy density between the first- and second-order phase transitions as 
\begin{align}
    \frac{\rho_\phi(T)}{\rho_\phi^{\mathrm{2nd}}(T)}
    =
    \frac{m_\mathrm{b}}{m_0} \cos^2 \alpha_\mathrm{b}
    +
    \frac{m_0}{m_\mathrm{b}} \sin^2 \alpha_\mathrm{b}
    \ ,
    \label{eq: axion ratio homogeneous mb > Hb}
\end{align} 
for $m_\mathrm{b} > 3H(T_\mathrm{b})$ and 
\begin{align}
    \frac{\rho_\phi(T)}{\rho_\phi^{\mathrm{2nd}}(T)}
    =
    \left( \frac{m_\mathrm{b}}{m_0} \right)^{\frac{6}{p}}
    \left( \frac{T_\mathrm{osc}}{\Lambda} \right)^{\frac{p+6}{2}}
    \ .
    \label{eq: axion ratio homogeneous mb < Hb}
\end{align} 
for $m_\mathrm{0} > 3H(T_\mathrm{b}) > m_\mathrm{b}$.

So far, for illustrative purposes, we have considered a setup where the axion mass change due to FOPT is assumed to occur uniformly in space. In the next section, we will consider the effect of inhomogeneity due to bubble nucleation. On the other hand, sudden changes in axion mass that occur uniformly in space, as considered here, may occur in other situations~\cite{Jeong:2022kdr}. For example, if the axion mass is determined by the dynamics of another scalar field, a similar axion mass change could be realized by a certain choice of the scalar potential.

\section{Bubble-induced axion wave dynamics}
\label{sec: axion wave dynamics}

\subsection{Classification of dynamics}
\label{subsec: classification}

So far, we have assumed that the FOPT takes place instantaneously and uniformly throughout the universe.
In realistic situations, however, the bubble wall expands with a finite velocity, and we need to take into account spatial inhomogeneities of $\phi$.
The time scale of the bubble expansion is given by $\beta^{-1}$, which is typically much shorter than the Hubble time at the bubble nucleation.
In the following, we consider the axion dynamics in four cases:
\begin{enumerate}
    \item[(a)] $3 H_\mathrm{b} > m_0$
    \item[(b)] $\beta > m_0 > m_\mathrm{b}, 3 H_\mathrm{b}$
    \item[(c)] $m_\mathrm{b} > \beta$
    \item[(d)] $m_0 > \beta > m_\mathrm{b}, 3 H_\mathrm{b}$
\end{enumerate}
{Here $\beta > H_\mathrm{b}$ and $m_0 > m_{\mathrm{b}}$ are assumed. 
Cases (b) and (d) are further classified based on the relative magnitude of $m_\mathrm{b}$ and $3H_\mathrm{b}$.}

In case (a), the axion starts to oscillate after the FOPT.
Thus, the axion abundance can be evaluated in the same way as in the standard misalignment mechanism with a constant mass $m_0$.
In case (b), the FOPT proceeds in a time scale shorter than the axion oscillation.
Then, we can apply the evaluation in the previous section;
we can use Eq.~\eqref{eq: axion ratio homogeneous mb > Hb} for $m_\mathrm{b} > 3 H_\mathrm{b}$ and Eq.~\eqref{eq: axion ratio homogeneous mb < Hb} for $m_\mathrm{b} < 3 H_\mathrm{b}$.
In cases (c) and (d), the finite velocity of the bubble wall significantly affects the evolution of the axion field in different ways.
In particular, the motion of bubble walls affects the axion field dynamics via two effects: the generation of axion waves and their reflection/transmission.

\subsection{Reflection and transmission}
\label{subsec: reflection & tranmission}

{%
The effect of the passage of the bubble wall on the {oscillating} axion field can be analyzed in terms of the reflection and transmission of axion waves by the bubble wall. In practice, it will be necessary to analyze axion waves among randomly generated spherical bubble walls, but the essence of the dynamics can be understood in terms of reflection and transmission by planar bubble walls.}
In the following, we assume that the width of the wall {in the wall rest frame} is significantly smaller than the de Broglie wavelength of the axion, allowing us to model the wall as a step function in its direction of motion.
Below we consider case (c), where the axion field has already started oscillating before FOPT.

Let us discuss the reflection and transmission of axion waves at a single bubble wall.
We show a schematic picture in Fig.~\ref{fig: ref & trans}.
We consider a bubble wall moving along the $z${-}axis and define the coordinate system $(t, z)$ such that the position of the wall is represented by $z = v t$.
To analyze the transmission and reflection of waves, it is most convenient to move to the rest frame of the bubble wall $(t', z')$ given by
\begin{align}
    t' &= \gamma (t - v z) \ ,
    \\
    z' &= \gamma (z - v t) \ ,
\end{align}
where $\gamma = 1/\sqrt{1 - v^2}$ is the Lorentz factor of the bubble wall, and the wall is located at $z'=0$.
We consider an axion wave propagating from the positive to the negative direction along the $z'$-axis in the wall frame in the form
\begin{align}
    \Phi_I= \phi_0 \exp [i (\omega' t'+ k_I' z' )]
    \ ,
\end{align}
where the real part of $\Phi$ corresponds to the axion field value.
Then, a fraction of this axion wave, say $\F_T$, penetrates the wall, and the rest, $\F_R$, is reflected. 
On the wall, at $z'=0$, we have the boundary conditions, 
\begin{align}
    \Phi_I[t',0] + \Phi_R[t',0] &= \Phi_T[t',0] 
    \ ,
    \\
    \partial_{z'}\Phi_I[t',0] + \partial_{z'}\Phi_R[t',0]
    &= \partial_{z'}\Phi_T[t',0]
    \ .
\end{align}
We then find the solution
\begin{align}
    \Phi_R[t',z']
    =
    \frac{k_I' - k_T'}{k_I' + k_T'} \phi_0 
    \exp [i (\omega' t'- k_R' z' )]
    \ ,
    \\
    \Phi_T[t',z']
    =
    \frac{2 k_I'}{k_I' + k_T'} \phi_0
    \exp [i (\omega' t'+ k_T' z' )]
    \ ,
\end{align}
with
\begin{align}
    k_R' = k_I'
    \ , \quad 
    k_T' 
    =
    \sqrt{k_I'^2 + m_\mathrm{b}^2 - m_0^2}
    \ .
\end{align}
This solution satisfies
\begin{align}
\label{conservation}
    k_I' |\Phi_I[t',0]|^2
    =
    k_R' |\Phi_R[t',0]|^2
    +
    k_T' |\Phi_T[t',0]|^2
    \ .
\end{align}
Noting that the axion number density is proportional to $\omega' |\Phi|^2$ and the group velocity is given by $k'/\omega'$, this equality implies the conservation of the axion number, which also holds in the cosmological frame.
\begin{figure}[!t]
    \begin{center}  
        \includegraphics[width=0.45\textwidth]{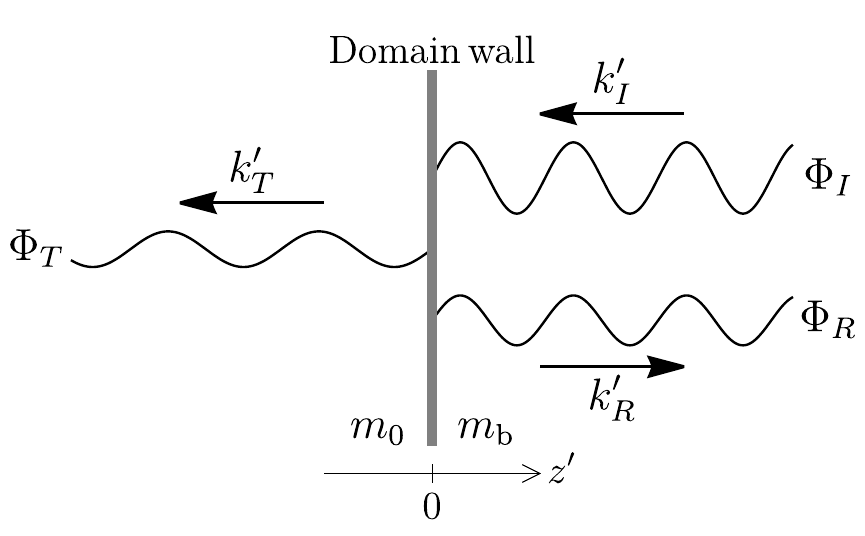}
        \end{center}
    \caption{%
        Reflection and transmission of the axion wave in the wall frame.
    }
    \label{fig: ref & trans} 
\end{figure}

In case (c), the axion field right before the bubble nucleation can be expressed as 
\begin{align}
    \Phi_I(t, z)
    =
    \bar{\phi}_\mathrm{b} \exp[i m_\mathrm{b} t]
    \ ,
\end{align}
in the cosmological frame.
Here, we ignored the cosmic expansion because the FOPT typically proceeds in a time scale much shorter than the Hubble time.
Then, moving to the wall frame, the axion becomes a wave propagating from the positive to the negative direction along the $z'$-axis in the form
\begin{align}
    \Phi_I(t', z')
    =
    \bar{\phi}_\mathrm{b} \exp [i (\omega_0' t'+ k_{I 0}' z' )]
    \ ,
\end{align}
with
\begin{align}
    \omega_0' = m_\mathrm{b} \gamma
    \ , \quad 
    k_{I 0}' = m_\mathrm{b} v \gamma
    \ .
\end{align}
Then, we obtain the momentum of the {transmitted} wave, $k_{T 0}'$, as 
\begin{align}
\label{kT0}
    {k_{T 0}'}
    =
    \sqrt{\gamma^2 m_\mathrm{b}^2 - m_0^2}
    \ .
\end{align}
When ${\rm Im}[k'_{T0}] \neq 0$, the field damps inside the bubble, and the amplitudes of $\F_R$ and $\F_I$ are the same except for a phase shift, indicating total reflection of the axion wave. The reflected wave is then repeatedly scattered off another bubble wall, and when it gains enough energy, it is transmitted into the bubble. This acceleration by repeated reflections is similar to the Fermi acceleration~\cite{Blandford:1978ky,Bell:1978clk,Drury:1983zz,Blandford:1987pw}, which will be studied next.

\subsection{Fermi acceleration}
\label{subsec: Fermi acceleration}

Let us continue with the $(1+1)$d spacetime simplification to describe the Fermi acceleration of the axion wave. We consider two walls moving towards each other at terminal velocity $v$. The wall moving in the positive direction along the $z$-axis is referred to as the ``up wall", and the other as the ``down wall". It is assumed that the axion wave continues to be completely reflected by the walls for a certain period.

Let us consider that a plain wave moves towards the up wall in the cosmological frame with a momentum $\{\omega, -k\}$. 
In the up wall frame, 
the axion wave has the four-momentum $\L(-v)\cdot \{\omega, -k\}$ with
\begin{align}
    \L(v)
    =
    \left( \begin{array}{cc}
        \gamma &v \gamma 
        \\ 
        v \gamma & \gamma 
    \end{array} \right)
    \equiv
    \left( \begin{array}{cc}
        \cosh \psi & \sinh \psi 
        \\ 
        \sinh \psi & \cosh \psi
    \end{array} \right)
    \ ,
\end{align}
where $\psi$ is the rapidity defined by $\tanh \psi = v$.
Then, this wave will be reflected, and the momentum becomes $\hat{P}\cdot\L(-v)\cdot \{\omega, -k\}$, where $\hat{P}=\diag{(1,-1)}$
denotes 
a parity transformation. 
In the cosmological frame, we get the momentum $\L(v)\cdot \hat{P}\cdot \L(-v)\cdot \{\omega,-k\}$. 
Then, this wave will be reflected by the down wall. 
With a similar transformation, we obtain the momentum in the cosmological frame after the second reflection as
$\L(-v) \cdot \hat{P} \cdot \L(v) \cdot \L(v) \cdot \hat{P} \cdot \L(-v) \cdot \{\omega, -k\}$.
By noting $\hat{P}\cdot \L(v)\cdot \hat{P}=\L(-v)$, we find the momentum in the cosmological frame after the $n$-th reflection 
\begin{align}
    p_\mu[n]
    &=
    \hat P^n \cdot \left(\L(-v)\right)^{2n} \cdot \{\omega,-k\}
    \nonumber \\
    &
    =
    \hat P^n \cdot 
    \begin{pmatrix}
        \cosh ( 2n\psi) & \sinh (-2n\psi)
        \\ 
        \sinh (-2n\psi) & \cosh ( 2n\psi)
    \end{pmatrix}
    \cdot \{\omega,-k\}
    . 
\end{align} 
Then, we get the angular frequency of the axion wave 
\begin{align}
    p_0[n]
    &=
        \frac{1}{2} \left[
        (\omega+k) \left(\frac{1+v}{1-v}\right)^n
        +
        (\omega-k) \left(\frac{1-v}{1+v}\right)^n
    \right]\ .
\end{align}
The enhancement of the angular frequency with $n\gg 1$ is given by the first term. 

The next question concerns the number of possible reflections. For simplicity, we assume that the axion wave propagates at the speed of light.%
\footnote{This implies that the first few reflections are neglected.}  
{Let us consider an axion wave that undergoes its first reflection at the up wall when the walls are separated by a distance $L$ and}
denote the distance between the walls at the $n$-th reflection as $L_{\rm w}[n]$. 
The initial condition is $L_{\rm w}[1] = L$. Then it takes a time of $L/(v +1)$ for the axion waves to reach the down wall and undergo the second reflection. The $v$ in the denominator takes into account the motion of the down wall. During this interval, the distance between the walls decreases by $2v L/(v +1),$ and thus the distance at the second reflection is $L_{\rm w}[2]=L(1-2v/(1+v))=\frac{1-v}{1+v} L$.  So at the $(n+1)$-th reflection we have 
$
L_{\rm w}[n+1]=\left(\frac{1-v}{1+v}\right)^n L.
$
This leads to 
\begin{align}
    n
    =
    \frac{\log{\frac{L}{L_{\rm w}[n+1]}}}
    {\log{\frac{1+v}{1-v}}}
    .
\end{align}
By using this expression and taking $\omega=m_\mathrm{b}$ and $k=0$, we find 
\begin{align}
    p_0[n]
     =
    \frac{m_\mathrm{b}}{2}
    \left(
    \frac{L}{L_{\rm w}[n+1]}+
        \frac{L_{\rm w}[n+1]}{L}
    \right)
    .
\end{align}
This enhancement continues with large $n$ until the approximation becomes invalid, for instance when transmission becomes important as we will discuss soon, or when the $L_{\rm w}$ becomes smaller than the wall width in the cosmological frame or the de Broglie wavelength of the axion waves.\footnote{The ratio of $L_{\rm w}$ to the de Broglie wavelength remains constant after the first scattering because of the Fermi acceleration.}

So far we have neglected transmission by assuming that {$k_T'$} is imaginary. However, reflection is not always perfect, especially when $n \gg 1$. For sufficiently large energies, {$k_T'$} becomes real, allowing transmission, which will be studied next.

\subsection{Transmission of axion waves}
\label{subsec: enhancement}
In the spatially uniform case, the number of the oscillating axion field increases when the axion mass in the FOPT suddenly increases. On the other hand,  when the effects of the bubble wall dynamics are taken into account, the axion number is conserved, as in Eq.~(\ref{conservation}).
This is one of the major changes caused by the inclusion of bubbles.
Below we will continue to consider case (c) and estimate how much of the axion wave is transmitted with each scattering.

As we have seen before, the transmission occurs when the angular frequency exceeds the mass inside the bubble:
\begin{align}
\label{cond}
    \omega_n' 
    > m_0,
\end{align}
where $\omega_n'$ is the angular frequency after the $n$-th reflection with the wall in the wall frame. 
Similarly to the previous discussion, we can estimate $\omega_n'$ as
\begin{align}
    \omega_n' &= a_n m_{\rm b}
    \ ,
    \\
    a_n &=
    \frac{1}{2} \left(
        \left(\sqrt{\frac{1+v}{1-v}}\right)^{2n-1} 
        +
        \left(\sqrt{\frac{1-v}{1+v}}\right)^{2n-1}
    \right)
    \ .
\end{align}

Let us assume that the inequality (\ref{cond}) is first met and the axion waves start to transmit when $n=n_c$.
We show the dependence of $n_c$ on $v$ and $m_0/m_\mathrm{b}$ in Fig.~\ref{fig: nc}.\footnote{Note that the cosmic expansion is neglected in this analysis, so the results for $v \ll 1$ must be interpreted with some caution. }
\begin{figure}[!t]
    \begin{center}  
        \includegraphics[width=0.45\textwidth]{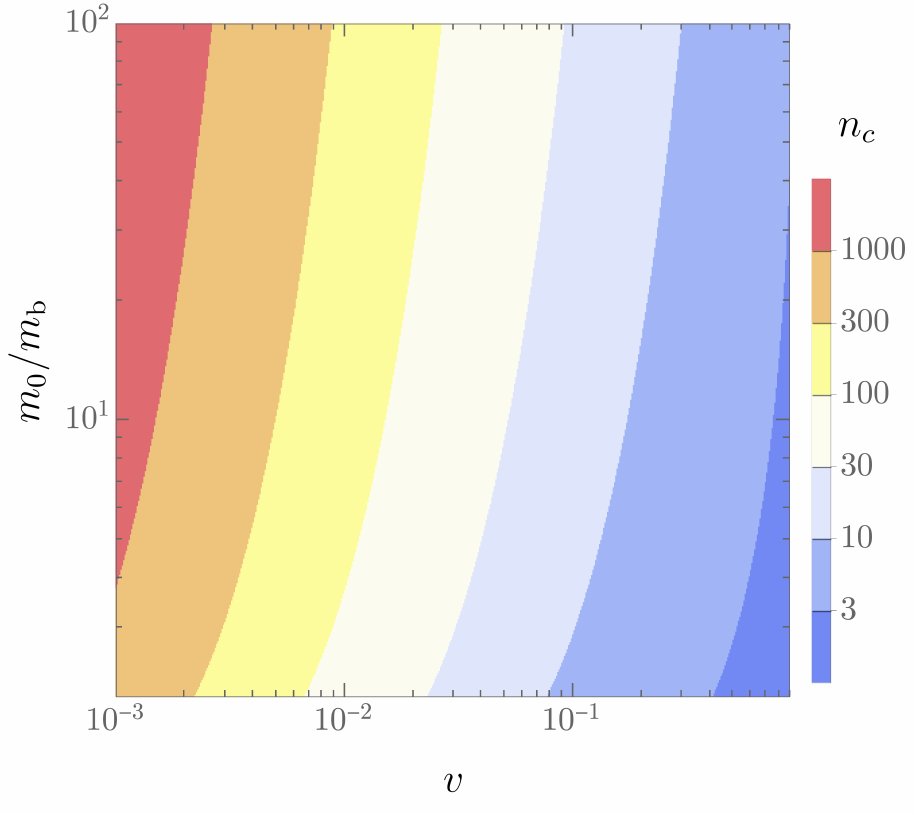}
        \end{center}
    \caption{%
        Parameter dependence of $n_c$.
        The smaller $v$ and the larger $m_0/m_\mathrm{b}$, the larger $n_c$ becomes.
    }
    \label{fig: nc} 
\end{figure}
The ratios of the axion flux of the transmitted and reflected waves to that of the incident wave at the $(n_c+i)$-th collision are given by $t_i$ and $r_i$ respectively, where $t_i$ and $r_i$ are defined by
\begin{align}
    t_i &= 
    \left|
        \frac{2 k_{I,n_c+i}'}{k_{I,n_c+i}'+k_{T,n_c+i}'}
    \right|^2 \times \frac{k_{T,n_c+i}'}{k_{I,n_c+i}'},
    \\
    r_i &= 
    \left|
        \frac{k_{I,n_c+i}'-k_{T,n_c+i}'}{k_{I,n_c+i}'+k_{T,n_c+i}'}
    \right|^2
\end{align}
with $k_{I,n}' \equiv \sqrt{\omega_n^{\prime 2} - m_{\rm b}^2}$ and $k_{T,n}' \equiv \sqrt{\omega_n^{\prime 2} - m_{0}^2}$.
Let $F_i$ be the fraction of axions transmitted at the $(n_c+i)$-th collision. It is defined as
\begin{align}
    F_0 \equiv t_0
    \ , \quad 
    F_i \equiv t_i r_0 \cdots r_{i-1} \quad (i \geq 1).
\end{align}
The sum of $F_i$ should approach unity.

We show $F_i$ in Fig.~\ref{fig: Fi}.
For $v=0.1$, most of the axion waves transmit within a few scatterings after $n = n_c$, and $F_i$ decreases rapidly with $i$. On the other hand, for $v \ll 1$, it takes more scatterings for the axion waves to transmit.
This is because, for larger $v$,  a larger fraction of the axion wave can transmit the bubble wall for small $i$ due to the larger acceleration at each collision with the wall.
\begin{figure}[!t]
    \begin{center}  
        \includegraphics[width=0.45\textwidth]{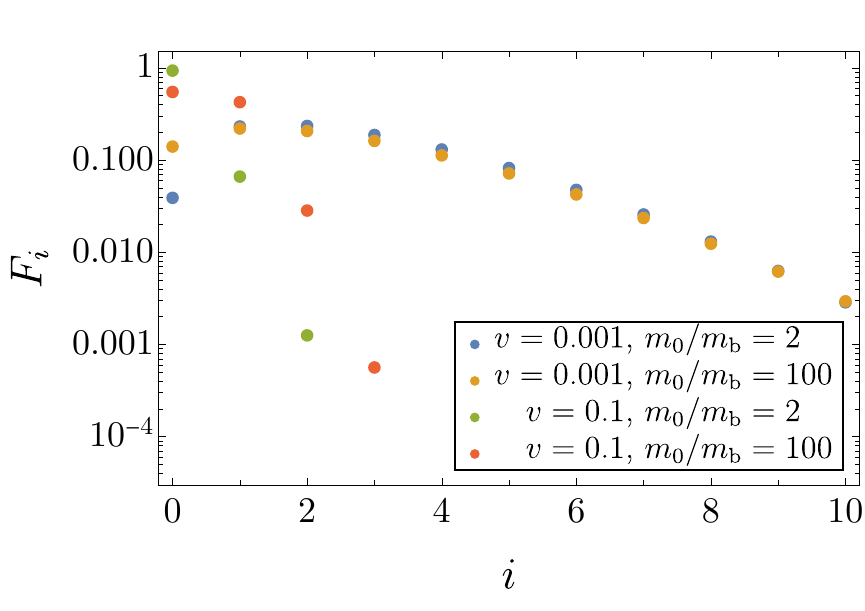}
        \end{center}
    \caption{%
        Fraction of axions transmitted at the $(n_c+i)$-th collision, $F_i$.
    }
    \label{fig: Fi} 
\end{figure}

The momentum distribution after the FOPT can also be estimated by boosting the transmitted momentum $k'_{T,n}$ with $n\geq n_c$ in the wall frame to $k_{T,n}$ in the cosmological frame. 
The momentum distribution is proportional to the fraction of the transmitted waves $\propto F_i$.
Thus, the typical momentum of the transmitted waves can be approximated by $k_{T,n_c}$, which we show in Fig.~\ref{fig: kT for nc}.
For $v<\mathcal{O}(0.1)$, $k_{T,n_c}$ is suppressed as $\sqrt{v}$.
On the other hand, for $v > \mathcal{O}(0.1)$, $k_{T,n_c}$ can take a range of values including negative values, which mean the axion wave propagating outward in the bubble.
This can be understood as follows.
Since axion waves are significantly accelerated at each reflection, $k_{T,n_c}$ tends to be large.
For some parameters, however, $\omega_{n_c}'$ is slightly larger than $m_0$, and $k_{T,n_c}'$ becomes much smaller than $m_0$.
Then, moving to the cosmological frame, we obtain negative $k_{T,n_c}$.\footnote{The wavenumbers in the wall frame are always positive by definition.}

Since the transmitted axion waves are typically non-relativistic or marginally non-relativistic, the spatial inhomogeneities of axion dark matter induced by FOPT {could} remain in the subsequent evolution of the universe.
\begin{figure}[!t]
    \begin{center}  
        \includegraphics[width=0.45\textwidth]{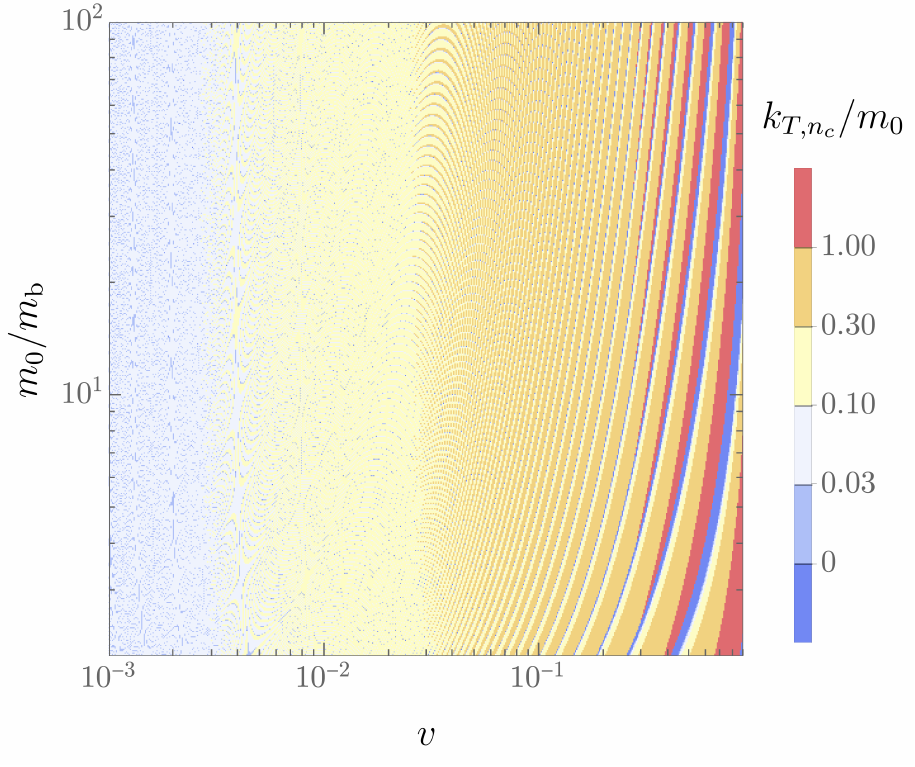}
        \end{center}
    \caption{%
        Momentum of the transmitted wave for $n = n_c$, $k_{T,n_c}$ as a function of $m_0/m_{\rm b}$ and $v$.
    }
    \label{fig: kT for nc} 
\end{figure}
The spatial scale of such fluctuations, $l$, can be estimated by the typical distance of the bubble walls when the $n_c$-th reflection occurs.
In particular, for $v \ll 1$, we obtain
\begin{align}
    l
    \simeq 
    R
    \left( \frac{1-v}{1+v} \right)^{n_c}
    \simeq 
    \frac{v}{\beta}
    \frac{m_\mathrm{b}}{2 m_0}
    \ ,
\end{align}
where we used
\begin{align}
    \omega_{n_c}'
    \simeq 
    \frac{1}{2} \left( \sqrt{\frac{1+v}{1-v}} \right)^{2n_c-1} m_\mathrm{b}
    \sim 
    m_0
    \ .
\end{align}
The axion free-streams over a distance of $(k_{T,n_c}/m_0) H_{\rm b}^{-1}$ during the Hubble time. If this is larger than $l$, the spatial inhomogeneities induced by the bubble wall dynamics are suppressed by the subsequent free streaming. Otherwise, the spatial inhomogeneities remain. This is the case if $\beta/H_{\rm b}$ is relatively small and the typical axion velocity is smaller than the wall velocity.

\subsection{Axion shock wave}
\label{subsec: axion shock wave}

In case (d), the axion does not oscillate outside the bubbles during the FOPT process. On the other hand, the axion mass inside the bubbles is large enough to affect the axion dynamics within the FOPT process.

First, we discuss analytically the axion dynamics just after bubble nucleation in case (d). We assume that the bubbles nucleate in a spherical shape and that the critical radius of the bubble is negligibly small compared to the cosmological scale. Then the axion field configuration becomes spherically symmetric and the equation of motion for the axion is given by 
\begin{align}
    \left[
        \frac{\partial^2}{\partial t^2}
        -\frac{1}{r^2} \frac{\partial}{\partial r}
        \left( r^2 \frac{\partial}{\partial r} \right)
        +
        m^2(t, r)
    \right]
    \phi(t,r)
    =
    0
    \ ,
\end{align}
which can be rewritten as 
\begin{align}
    \left[
        \frac{\partial^2}{\partial t^2}
        -\frac{\partial^2}{\partial r^2}
        +
        m^2(t, r)
    \right]
    u(t,r)
    =
    0
    \ ,
\end{align}
Here $r$ is the radius, and we define $u \equiv r \phi$, and $m^2$ is the axion mass.
In the following, we redefine the origin of the time coordinate so that $t = 0$ corresponds to the bubble nucleation.
Then, $m^2$ is given by 
\begin{align}
    m^2(t, r)
    =
    \left\{
        \begin{array}{ll}
            m_0^2 
            & \quad
            (0 \leq r < vt)
            \\
            m_\mathrm{b}^2
            & \quad
            (r > v t)
        \end{array}
    \right.
    \ .
\end{align} 
Since $\beta > m_\mathrm{b}$, we can neglect $m_\mathrm{b}$ and approximate the initial condition as
\begin{equation}
\begin{aligned}
    \phi(t = 0, r)
    &= 
    \phi_\mathrm{b}
    \ , 
    \\
    \frac{\partial \phi}{\partial t}(t = 0, r)
    &=
    0
    \ .
\end{aligned}
\end{equation}
In particular, if $m_\mathrm{b} < 3H_\mathrm{b}$, we can use $\phi_\mathrm{b} = \phi_0$.

As the bubble expands, the axion inside the bubble evolves with a mass of $m_0$.
If the bubble velocity is sufficiently smaller than the speed of light, the axion field near inside the bubble wall does not oscillate due to the gradient energy around the bubble wall.
Then, the axion field inside the bubble settles to zero due to the mass.
The depletion of the axion field value inside the bubble propagates with the speed of light to outside the bubbles, and its effect decreases with distance.
Consequently, we can approximate that the axion field changes its value from $0$ at $r = vt$ to $\phi_\mathrm{b}$ at $r = t$.
By using these boundary conditions, we can solve the axion field configuration in $vt < r < t$ and obtain an approximate solution as~\footnote{%
Note that similar approximate solutions can be derived in 2-dimensional space, but this argument does not apply to 1-dimensional space.
}
\begin{align}
    \phi(t, r)
    \simeq 
    \left\{
        \begin{array}{ll}
            0 
            & \quad
            (0 \leq r < vt)
            \\
            \left[ 
                1 - \frac{v (t - r)}{(1 - v) r}
            \right] \phi_\mathrm{b}
            & \quad
            (vt \leq r < t)
            \\
            \phi_\mathrm{b}
            & \quad
            (r \geq t)
        \end{array}
    \right.
    \ .
\end{align} 
We call this axion field configuration associated with a bubble an axion shock wave.
We show in Fig.~\ref{fig: Fitting 3d} the numerical result and the approximate solution, and it can be seen that they are in good agreement with each other. 
\begin{figure}[!t]
    \begin{center}  
        \includegraphics[width=0.45\textwidth]{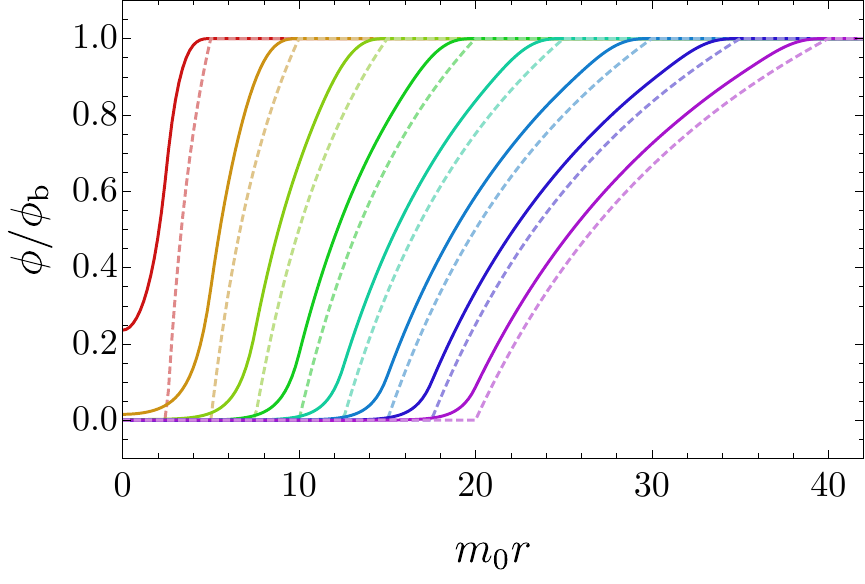}
        \end{center}
    \caption{%
        Evolution of the axion field in case (d) with  $v = 0.5$. The solid and dashed lines represent the numerical result and the approximate solution, respectively.   The results are shown for $5 \leq m_0 t \leq 40$ at intervals of $m_0 \Delta t = 5$.
        The axion is depleted inside the bubble.
    }
    \label{fig: Fitting 3d} 
\end{figure}

We estimate the axion number density after the FOPT.
The energy of the radiated axion wave is given by 
\begin{align}
    E_\phi
    &\equiv 
    4\pi \int_{vt}^{t}
    \mathrm{d}r \, 
    r^2 \frac{(\partial_t \phi)^2 + (\partial_r \phi)^2}{2}
    \nonumber \\
    &\simeq
    2 \pi \frac{ v (1+v)}{1-v} \phi_\mathrm{b}^2 t
    \ .
\end{align}
The typical momentum of the axion field is given by 
\begin{align}
    k_\mathrm{typ}
    &\equiv
    \left. 
        \frac{\partial_r \phi}{\phi}
    \right|_{r = (1+v)t/2}
    \simeq
    \frac{4 v}{(1-v^2)t}
    \ .
\end{align}
Then, the axion number associated with a single bubble before collisions is estimated by 
\begin{align}
    N_\phi
    \equiv 
    \frac{E_\phi}{k_\mathrm{typ}}
    \simeq
    \frac{\pi \phi_\mathrm{b}^2}{2}
    (1+v)^2 t^2
    \ .
\end{align}
Thus, the axion number increases with time, implying that the axion wave continues to be produced as the bubble expands.

Since the axion number is conserved in reflections and transmissions with the bubble wall, 
we obtain the axion number density after the FOPT as
\begin{align}
    n_{\phi}
    \equiv 
    \frac{N_{\phi}}{L^3}
    \simeq
    \frac{\pi \phi_\mathrm{b}^2}{2 L}
    \ ,
    \label{nphi}
\end{align}
where we assumed {$t = L/(1+v)$ as a typical time when the axion wave first scatters off the bubble wall} and neglected the velocity dependence because it depends sensitively on how the typical momentum is chosen. 
It is very important to note that the density of axion numbers is inversely proportional to the distance $L$ between bubbles. This indicates that as the distance between bubbles increases, the production of axion number decreases. This phenomenon occurs because in case (d) the axion waves are restricted to the region just outside the bubbles, in contrast to case (c), where the axion waves are spread over the entire space from the beginning.

When the axion shock wave scatters off another bubble wall, it gets accelerated similarly to the previous case. Thus, the Fermi acceleration and the transmission

We will confirm these observations in numerical simulations next.

\subsection{Numerical results}
\label{subsec: numerical result}
Due to the complexity of the dynamics of axion waves, numerical lattice calculation is essential to study them in a more realistic setting than those modeled by planar walls or spherical axion waves. Here, for simplicity, we focus on case (d) with $m_{\rm b}=0$, neglecting the cosmic expansion, and perform numerical lattice calculations to confirm that the analytical estimate well describes the axion abundance in the final stage of the transition. At the end of this subsection, we will briefly comment on case (c).

In the three-dimensional lattice simulations, we have set a bubble nucleated at the center of the lattice {box} and imposed the periodic boundary condition.
This implies that bubble nucleation points are aligned with the equal separations determined by the lattice box size $L_{\rm box}$, which corresponds to $v/\beta$.
Here the expansion of the universe and the mass outside the bubble are neglected, assuming $\beta \gg m_{\rm b}, 3H_{\rm b}$.

The equation of motion is given by
\begin{equation}
    \left[ \frac{\partial^2}{\partial t^2} - \bm{\nabla}^2 + m^2(t,\bm{x}) \right] \phi(t,\bm{x}) = 0.
\end{equation}
Here, the mass squared of the axion is set to be
\begin{equation}
    m^2(t,\bm{x}) = \frac{1-\tanh{((r-vt)/\delta_w)}}{2} \times m_0^2,
\end{equation}
where $r$ is the distance from the bubble nucleation point and $\delta_w$ determines the thickness of the bubble wall in the rest frame of the bubble nucleation point, not in the wall rest frame.
In all simulations, we choose $\delta_w = 0.15 m_0^{-1}$, which is smaller than the typical de Broglie wavelength of the axion wave.
We have checked that thinning the wall does not significantly change the results.

The parameter settings for the simulations are summarized in Table.\,\ref{tab: params_3Dlat}.
Each simulation starts with the homogeneous initial condition $\phi = \phi_0$ and
ends at $t_{\rm end} = L_{\rm box}/v$, sufficiently later than the end of the phase transition.

We show an example of the time evolution of the axion's energy density in Fig.~\ref{fig: energy_density_time_evo}.
The solid line is the total energy density of the axion, while the {dotted} line is the contribution of the bubbles, the energy inside the bubbles divided by the total volume.
Each bounce corresponds to the reflection and transmission of the axion waves, so the interval decreases as the bubble wall approaches the nearest wall. The Fermi acceleration appears as an increase in the energy density outside the bubble, and as the energy increases, the transmission into the bubble becomes more prominent.

Through simulations, we have calculated the axion number density after the end of the phase transition, $n_{\phi}$, which is given by
\begin{equation}
    n_{\phi} = \frac{1}{V}\int \frac{{\rm d}^3 \bm{k}}{(2\pi)^3 2\omega_k} \left[ \dot{\tilde{\phi}}^2 + \omega_k^2 \tilde{\phi}^2 \right] ,
\end{equation}
where $\tilde{\phi}(t,\bm{k})$ is the Fourier transform of $\phi(t,\bm{x})$, $V = L_{\rm box}^3$ is the volume of the system, and $\omega_k \equiv \sqrt{\bm{k}^2 + {m_0^2}}$.
We show in Fig.~\ref{fig: number_density_vel_int}  the numerical results of the axion's number density $n_{\phi}$ for the box size $L_{\rm box}$.
The circular, square, and triangular points correspond to the results of $v=0.1\, , 0.5$, and $0.9$ respectively.
The black dotted line is proportional to $L_{\rm box}^{-1}$.
We can observe the excellent agreement with the power law expected in Eq.~(\ref{nphi}).
\begin{table}[!t]
    \centering
    \begin{tabular}{>{\centering\arraybackslash}p{0.085\textwidth} >{\centering\arraybackslash}p{0.085\textwidth} >{\centering\arraybackslash}p{0.085\textwidth} >{\centering\arraybackslash}p{0.085\textwidth} >{\centering\arraybackslash}p{0.085\textwidth}}
        \hline\hline
        Case & $(m_0L_{\rm box})^3$ & ${N_{\rm grid}}^3$ & $v$ & $m_0t_{\rm end}$ \\
        \hline
        (v1-1)  & $(2\pi)^3$  & $128^3$  & 0.1 & 62.8   \\
        (v1-2)  & $(4\pi)^3$  & $256^3$  & 0.1 & 126.7  \\
        (v1-3)  & $(8\pi)^3$  & $512^3$  & 0.1 & 251.3  \\
        (v1-4)  & $(16\pi)^3$ & $1024^3$ & 0.1 & 502.7  \\
        (v5-1)  & $(2\pi)^3$  & $128^3$  & 0.5 & 12.6   \\
        (v5-2)  & $(4\pi)^3$  & $256^3$  & 0.5 & 25.1   \\
        (v5-3)  & $(8\pi)^3$  & $512^3$  & 0.5 & 50.3   \\
        (v5-4)  & $(16\pi)^3$ & $1024^3$ & 0.5 & 100.5  \\
        (v9-1)  & $(2\pi)^3$  & $128^3$  & 0.9 & 7.0    \\
        (v9-2)  & $(4\pi)^3$  & $256^3$  & 0.9 & 14.0   \\
        (v9-3)  & $(8\pi)^3$  & $512^3$  & 0.9 & 27.9   \\
        (v9-4)  & $(16\pi)^3$ & $1024^3$ & 0.9 & 55.9   \\
        \hline\hline
    \end{tabular}
    \caption{%
    The parameter setup for the 3D lattice simulations.
    }
    \label{tab: params_3Dlat}
\end{table}
\begin{figure}[!t]
    \begin{center}  
        \includegraphics[width=0.425\textwidth]{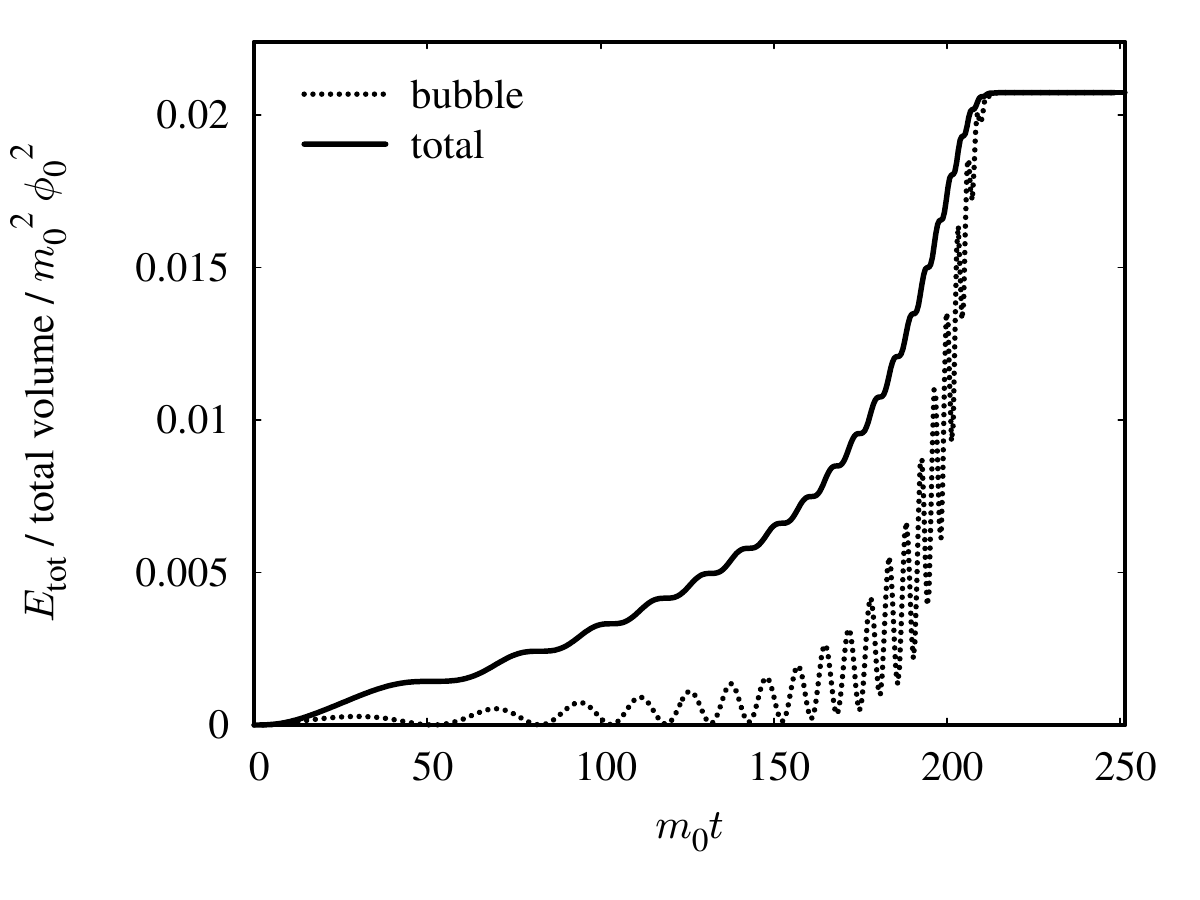}
    \end{center}
    \caption{%
    The time evolution of the energy of the system in the case (v1-3) where $L_{\rm box} = 8\pi m_0^{-1}$ and $v=0.1$.
    The solid line represents the total average energy density, whereas
    the dotted line is the energy density inside the bubble times the volume fraction of the bubble.    
    }
    \label{fig: energy_density_time_evo}
\end{figure}

\begin{figure}[!t]
    \begin{center}  
        \includegraphics[width=0.425\textwidth]{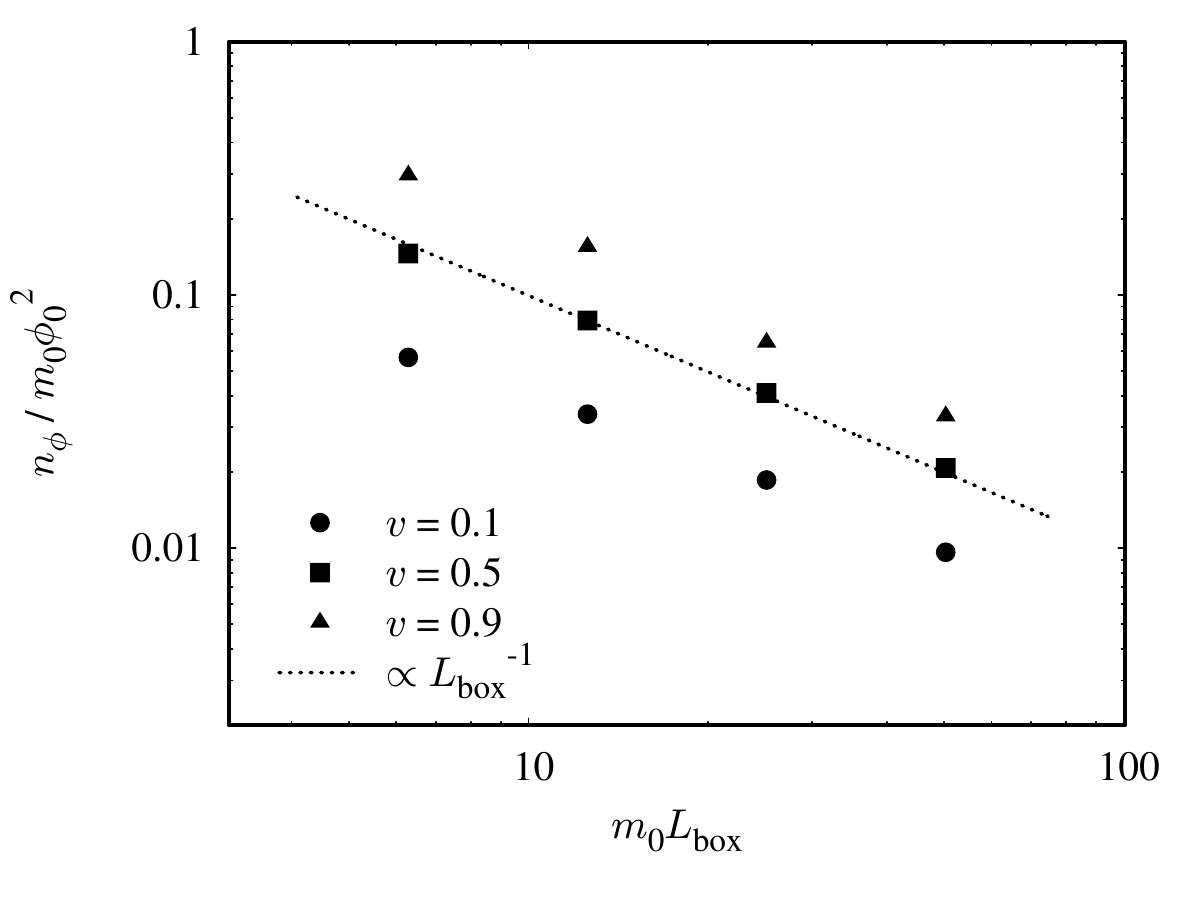}
    \end{center}
    \caption{%
    The axion number density at the final time as a function of the box size for $v = 0.1, 0.5$ and $0.9$.
    The dotted line is the line proportional to $L_{\rm box}^{-1}$.
    }
    \label{fig: number_density_vel_int}
\end{figure}

Since the wall velocity dependence of the final number density does not vary significantly for different $L_{\rm box}$ in Fig.~\ref{fig: number_density_vel_int}, we have studied the axion number density for different wall velocities, fixing $L_{\rm box} = 16\pi m_0^{-1}$ and $N_{\rm grid} = 1024$.
The results are shown in Fig.~\ref{fig: number_density_Lbox_int}.
The number density behaves differently in the high-velocity region and the low-velocity region pivoting $v \sim 0.4$.
Two dashed lines represent the result of piecewise power law fits.
The fitted results are given by
\begin{equation}
    n_{\phi} = C v^{\alpha} \frac{\phi_0^2}{L_{\rm box}} ,
\end{equation}
with
\begin{equation}
    C \simeq
    \left\{
        \begin{array}{ll}
             1.3 & \quad (v \lesssim 0.4) \\
             1.8 & \quad (v \gtrsim 0.4)
        \end{array}
    \right.
    ,
    \quad
    \alpha \simeq
    \left\{
        \begin{array}{ll}
             0.4 & \quad (v \lesssim 0.4) \\
             0.8 & \quad (v \gtrsim 0.4)
        \end{array}
    \right.
    .
\end{equation}
Thus, these numerical results agree well with the analytical estimate in Eq.~(\ref{nphi}).

We show in Fig.~\ref{fig: occupation_spectrum_int} the momentum distribution of the created axion number, $n(k)$, at the final time of each simulation where the box size is fixed, $L_{\rm box} = 16\pi m_0^{-1}$.
Here $n(k)$ is defined by
\begin{equation}
    n(k) = \frac{k^3}{2\pi^2 \cdot 2\omega_k V}\left[ \dot{\tilde{\phi}}^2 + \omega_k^2 \tilde{\phi}^2 \right] ,
\end{equation}
where $k\equiv |{\bm{k}}|$ assuming the rotational symmetry, so that $n_{\phi} = \int {\rm d}\ln{k}\, n(k)$.
We can see that the generated axions are marginally non-relativistic, and that the peak momentum of the axion increases with the wall velocity, which makes sense since the axion gains more energy with each reflection.

\begin{figure}[!t]
    \begin{center}  
        \includegraphics[width=0.425\textwidth]{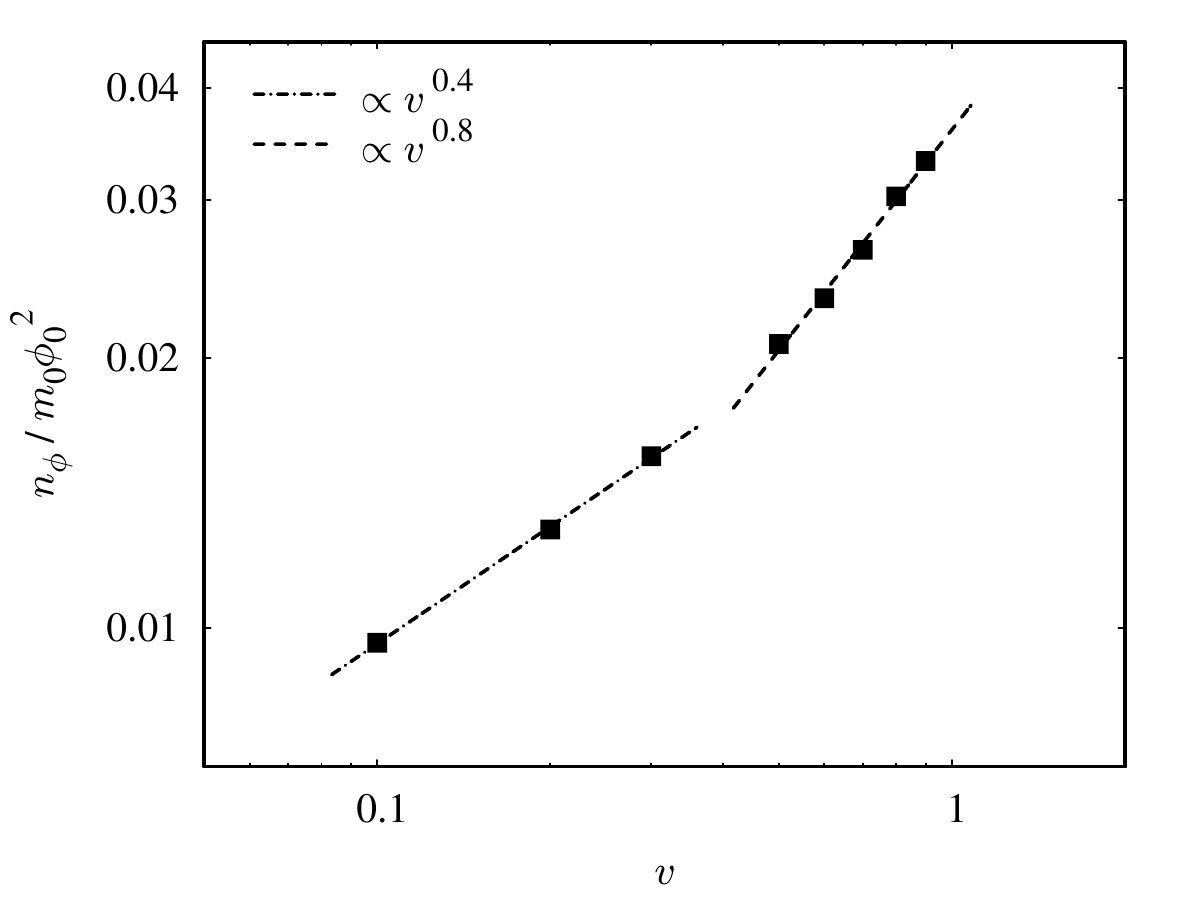}
    \end{center}
    \caption{%
   {The axion number density at the final time as a function of the bubble velocity.}
    }
    \label{fig: number_density_Lbox_int}
\end{figure}

\begin{figure}[!t]
    \begin{center}  
        \includegraphics[width=0.425\textwidth]{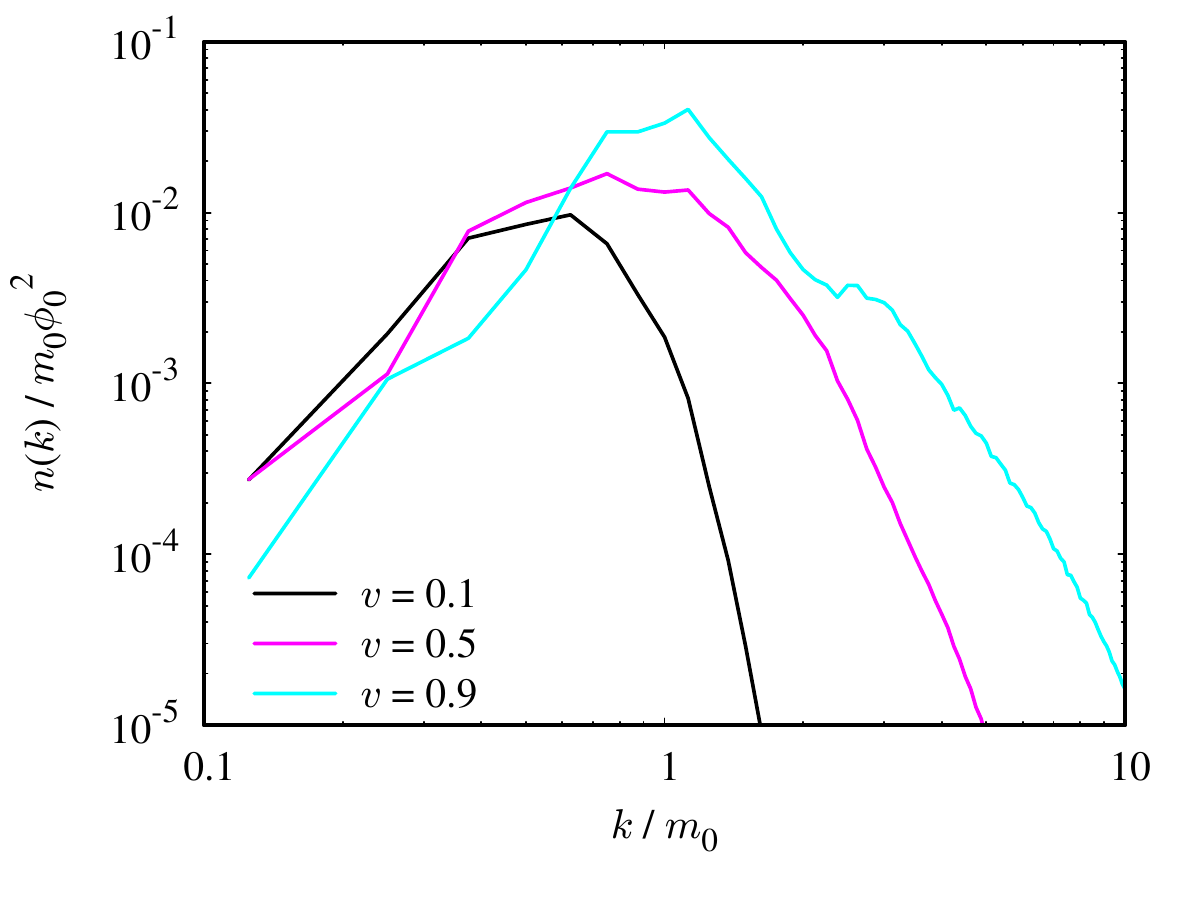}
    \end{center}
    \caption{%
    The momentum distribution of the axion number density after the end of the phase transition.
    Three lines correspond to the result with the wall velocity $v=0.1$ (black), $0.5$ (magenta), and $0.9$ (cyan), and we set $L_{\rm box} = 16\pi m_0^{-1}$.
    }
    \label{fig: occupation_spectrum_int}
\end{figure}

Lastly, we comment on the numerical results of case (c), in which the axion number is preserved during the bubble collision process while the energy density is enhanced via Fermi acceleration. Note that, while there is a single bubble in the lattice box, there are effectively multiple bubbles due to the periodic boundary condition.
To take a sufficiently spacious box size for the axion outside bubbles to oscillate until the bubble walls collide, i.e., $L/(2v) \gg m_{\rm b}^{-1}$, we set the mass squared of the axion to be a step function rather than a hyperbolic tangent. We have confirmed that this change of the bubble wall width does not change our results.
The triangular points in Fig.\,\ref{fig: number_density_caseC} show the number density after all the walls disappear.
The lattice with ${N_{\rm grid}}^3 = 512^3$ and $(m_0L_{\rm box})^3 = (16\pi)^3$ is used and $m_0/m_{\rm b} = 2$ is chosen. For this choice of the parameters, the initial axion number density is given by $n_{\phi}/m_0 \phi_0^2 = \frac{1}{2}m_{\rm b}/m_0 = 0.25$.
The numerical result implies that the axion number is conserved robustly, independent of $v$.
This should be contrasted to case (d) shown in Fig.~\ref{fig: number_density_Lbox_int}.
In contrast, due to the Fermi acceleration, the final energy density increases with wall velocity.
\begin{figure}[!t]
    \begin{center}  
        \includegraphics[width=0.425\textwidth]{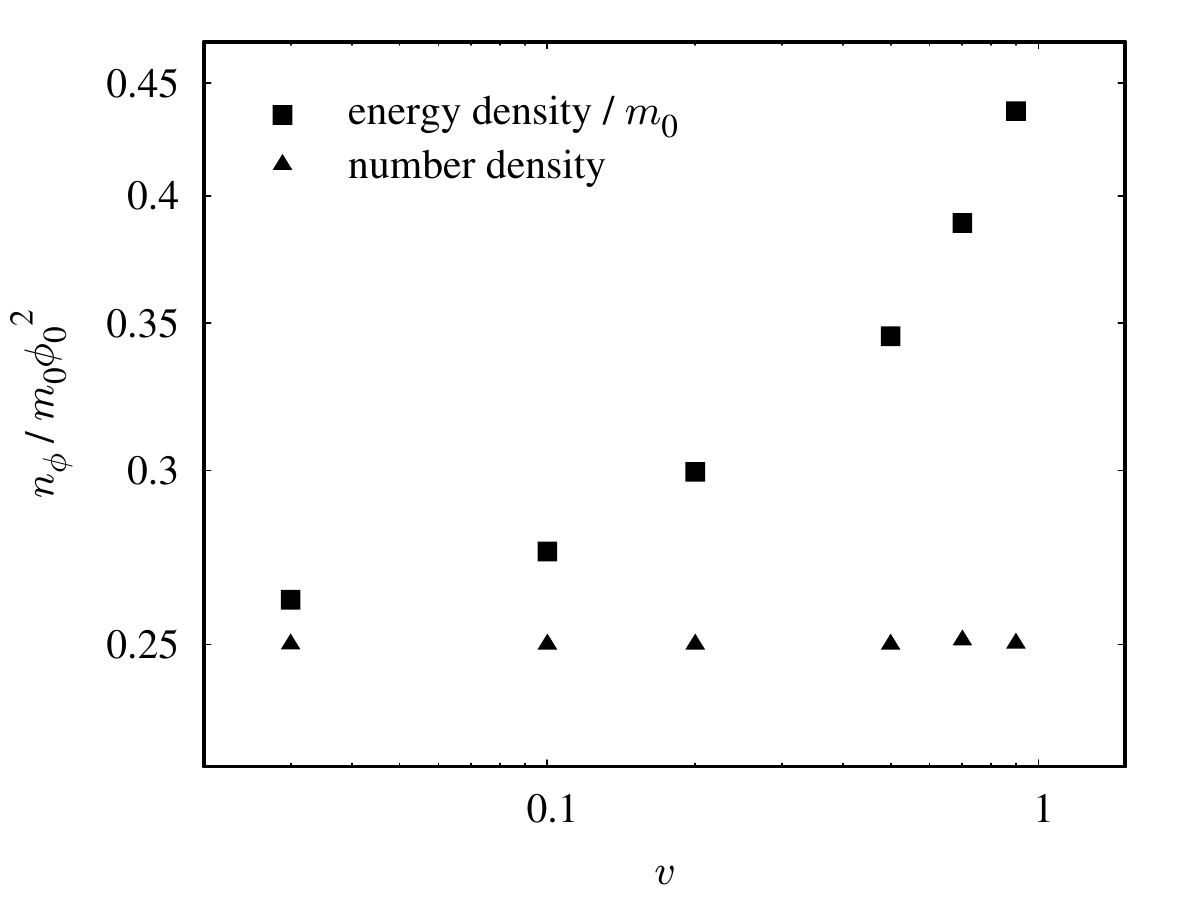}
    \end{center}
    \caption{%
    The axion number density and the energy density at the final time as a function of the bubble velocity.
    }
    \label{fig: number_density_caseC}
\end{figure}

\section{Bubble misalignment mechanism}
\label{sec: bubble misalignment}

\subsection{Scenario overview}
\label{subsec: scenario overview}
Before calculating the axion abundance, let us first discuss the axion dynamics in the four cases we have classified, based on the results of the previous section. For readability, the four cases are noted here again: (a) $3 H_\mathrm{b} > m_0$; (b) $\beta > m_0 > m_\mathrm{b}, 3 H_\mathrm{b}$; (c) $m_\mathrm{b} > \beta$; (d) $m_0 > \beta > m_\mathrm{b}, 3 H_\mathrm{b}$.

In case (a), the axion does not oscillate even after the FOPT is completed but starts to oscillate after a while. The misalignment mechanism works as in the case where the axion mass has a constant value of $m_0$.

On the other hand, in case (b) the axion does not move during the FOPT, but starts oscillating with the mass $m_0$ immediately afterwards.  For $m_\mathrm{b} < 3H(T_\mathrm{b})$, the axion did not start oscillating before the FOPT. Let us call this case (b1). This situation corresponds to that studied in Ref.~\cite{Nakagawa:2022wwm} and is similar to the trapped misalignment mechanism~\cite{Higaki:2016yqk,DiLuzio:2021gos,Jeong:2022kdr}. For $m_\mathrm{b} > 3H(T_\mathrm{b})$, the axion has already started oscillating before the FOPT. Let us call this case (b2). In cases (a) and (b), bubbles do not significantly affect the axion dynamics.

The bubble plays an interesting and crucial role in cases (c) and (d). In case (c), the axion has already started oscillating before the FOPT, and it oscillates outside the bubbles even during the FOPT.
The nucleated bubbles expel axions from the inside if $m_0 > \gamma m_\mathrm{b}$ (see Eq.~\eqref{kT0}). These axion waves outside the bubbles repeatedly scatter off the bubble walls until they acquire enough energy to enter the bubbles. In this Fermi acceleration process, the axion number is conserved.

In case (d), the axion does not oscillate outside the bubbles during the FOPT,
and it is further classified into cases (d1) and (d2) for $m_\mathrm{b} < 3H(T_\mathrm{b})$ and $m_\mathrm{b} > 3H(T_\mathrm{b})$, respectively. In case (d), as the bubbles nucleate, the axions are expelled from the inside of the bubbles, similar to case (c). The difference is that outside the bubbles, the axion forms a kind of shock wave, and their momentum decreases as they propagate. Accordingly, the effective number of axions increases over time until they finally enter the bubbles after acquiring enough energy through Fermi acceleration. In contrast to case (c), the resultant axion abundance is a decreasing function of the typical distance between the bubbles because the axion shock wave exists only near the surface of the bubbles. However, in case (c), the axion is already oscillating throughout space before the FOPT. 

\subsection{Axion abundance}
\label{subsec: axion abundance}

Here, we summarize the predictions for axion abundance in each case. For simplicity, we assume that the universe is radiation-dominated, where all components in the plasma have the same temperature. It is straightforward to estimate axion abundance during a matter-dominated era, such as the inflaton oscillation period, or in cases where the hidden gauge sector has a different temperature from the standard model plasma.

First, as a reference, we evaluate the axion abundance in the second-order phase transition.
Assuming the topological susceptibility given in Eq.~\eqref{eq: chi}, we obtain 
\begin{align}
    \frac{\rho_\phi^\mathrm{(2nd)}}{s}
    &=
    \frac{
        \frac{1}{2} m_0 m_\phi(T_\mathrm{osc}) \phi_0^2
    }
    {\frac{2 \pi^2 g_{*s,\mathrm{osc}}}{45}T_\mathrm{osc}^3}
    \nonumber \\
    &=
    \frac{45 }{4 \pi^2 g_{*s,\mathrm{osc}}}
    \left( \frac{\pi^2 g_{*,\mathrm{osc}}}{10} \right)^{\frac{p+6}{2(p+4)}}
    \frac{m_0^{\frac{p+2}{p+4}} \phi_0^2}{M_\mathrm{Pl}^{\frac{p+6}{p+4}} \Lambda^{\frac{p}{p+4}}}
    \nonumber \\
    &=
    \frac{45 }{4 \pi^2 g_{*s,\mathrm{osc}}}
    \left( \frac{\pi^2 g_{*,\mathrm{osc}}}{10} \right)^{\frac{p+6}{2(p+4)}}
    \frac{m_0 \phi_0^2}{(m_\mathrm{b}^2 M_\mathrm{Pl}^{p+6} T_\mathrm{b}^p)^\frac{1}{p+4}}
    \ .
    \label{rho2ssec}
\end{align}
Here and in the following, $g_{*s,\mathrm{osc}}$ represents the effective number of relativistic species contributing to the entropy, evaluated at the moment when the axion first begins to oscillate.

In case (a), the axion does not start oscillating yet at FOPT, but starts oscillating at $m_0 \sim 3 H$. The temperature at the onset of oscillations is given by 
\begin{align}
    T_\mathrm{osc,(a)}
    \simeq 
    \left( \frac{10}{\pi^2 g_{*,\mathrm{osc}}} \right)^{1/4}
    \sqrt{M_\mathrm{Pl} m_0}
    \ .
\end{align}
Then, we obtain
\begin{align}
    \frac{\rho_\phi^\mathrm{(a)}}{s}
    &=
    \frac{\frac{1}{2}m_0^2 \phi_0^2}
    {\frac{2 \pi^2 g_{*s,\mathrm{osc}}}{45}T_\mathrm{osc,(a)}^3}
    \nonumber \\
    &=
    \frac{45}{4\pi^2 g_{*s,\mathrm{osc}}}
    \left( \frac{\pi^2 g_{*,\mathrm{osc}}}{10} \right)^{3/4}
    \frac{m_0^{1/2} \phi_0^2}{M_\mathrm{Pl}^{3/2}}
    \ .
\end{align}
This is of course equal to the axion abundance when the mass of the axion is constant at $m_0$. 

In case (b), from the discussion in Sec.~\ref{sec: spatially uniform}, we obtain
\begin{align}
    \frac{\rho_\phi^\mathrm{(b1)}}{s}
    &=
    \frac{\frac{m_0^2 \phi_0^2}{2}}
    {\frac{2 \pi^2 g_{*s,\mathrm{osc}}}{45}T_\mathrm{b}^3}
    \nonumber \\
    &=    
    \frac{45 m_0^2 \phi_0^2}
    {4 \pi^2 g_{*s,\mathrm{osc}}T_\mathrm{b}^3}
    \ ,
\end{align}
for $m_\mathrm{b} < 3H(T_\mathrm{b})$ and 
\begin{align}
    \frac{\rho_\phi^\mathrm{(b2)}}{s}
    &=
    \frac{
        \frac{\phi_0^2}{2}
        \frac{m_\phi(T_\mathrm{osc,(b2)})}{m_\mathrm{b}}
    }
    {\frac{2 \pi^2 g_{*s,\mathrm{osc}}}{45}T_\mathrm{osc,(b2)}^3}
    \left( 
        m_\mathrm{b}^2 \cos^2 \alpha_\mathrm{b}
        +
        m_0^2 \sin^2 \alpha_\mathrm{b}
    \right)
    \nonumber \\
    &=
    \frac{45}
    {4 \pi^2 g_{*s,\mathrm{osc}}}
    \left( 
        \frac{\pi^2 g_{*,\mathrm{osc}}}{10}
    \right)^{\frac{p+6}{2(p+4)}}
    \frac{\phi_0^2}
    { (m_\mathrm{b}^{p+6} M_\mathrm{Pl}^{p+6}
    T_\mathrm{b}^p)^{\frac{1}{p+4}}}
    \nonumber \\
    &\phantom{=}\times
    \left( 
        m_\mathrm{b}^2 \cos^2 \alpha_\mathrm{b}
        +
        m_0^2 \sin^2 \alpha_\mathrm{b}
    \right)
    \ ,
\end{align}
for $m_\mathrm{b} > 3H(T_\mathrm{b})$.

In case (c), the axion number is conserved by the reflection and transmission processes. Thus the axion abundance is the same as in the case of the second-order phase transition given in Eq.~(\ref{rho2ssec}).
Thus, we obtain
\begin{align}
    \frac{\rho_\phi^\mathrm{(c)}}{s}
    &=
     \frac{45 }{4 \pi^2 g_{*s,\mathrm{osc}}}
    \left( \frac{\pi^2 g_{*,\mathrm{osc}}}{10} \right)^{\frac{p+6}{2(p+4)}}
    \frac{m_0 \phi_0^2}{(m_\mathrm{b}^2 M_\mathrm{Pl}^{p+6} T_\mathrm{b}^p)^\frac{1}{p+4}}
    \ .
\end{align}

Case (d) can be further classified into case
(d1) for $m_\mathrm{b} < 3H(T_\mathrm{b})$ and case (d2) for $m_\mathrm{b} > 3H(T_\mathrm{b})$.
We estimate the axion abundance in case (d1) using the numerical results.
In the numerical simulations, the adjacent bubble is separated by a distance of $L_\mathrm{box}$.
Thus, we estimate the axion number density just after the FOPT as
\begin{align}
    n_\phi(T_\mathrm{b}) 
    \simeq 
    C v^{-1+\alpha} \beta \phi_0^2
\end{align}
by setting $L_\mathrm{box} = v/\beta$.
Then, we obtain the axion energy density in the later universe as
\begin{align}
    \frac{\rho_\phi^\mathrm{(d1)}}{s}
    &\simeq
    \frac{C v^{-1+\alpha} \beta m_0 \phi_0^2}
    {\frac{2 \pi^2 g_{*s,\mathrm{b}}}{45}T_\mathrm{b}^3}
    \nonumber \\
    &=
    \frac{45 C v^{-1+\alpha} \beta m_0 \phi_0^2}
    {2 \pi^2 g_{*s,\mathrm{b}} T_\mathrm{b}^3}
    \ ,
\end{align}
where $g_{*s,\mathrm{b}} \equiv g_{*s}(T_\mathrm{b})$.

On the other hand, in case (d2), the axion field at the FOPT has a degree of freedom of $\alpha_\mathrm{b}$ as in case (b2). $\phi$ and $\dot{\phi}$ are uniform in space during the timescale of $1/\beta (\ll 1/m_{\rm b})$.
If $\dot{\phi} = 0$ at the FOPT, the axion dynamics during the FOPT process will be the same as in case (d1), and the axion abundance is given by
\begin{align}
    \frac{\rho_\phi^\mathrm{(d2\text{-}1)}}{s}
    &\simeq
    \frac{C v^{-1+\alpha} \beta m_0 \bar{\phi}_\mathrm{b}^2}
    {\frac{2 \pi^2 g_{*s,\mathrm{b}}}{45}T_\mathrm{b}^3}
    \nonumber \\
    &=
    \frac{45 C v^{-1+\alpha} \beta m_0 \phi_0^2}
    {2 \pi^2 g_{*s,\mathrm{b}} T_\mathrm{b}^3}
    \left( \frac{T_\mathrm{b}}{T_\mathrm{osc}} \right)^{(p+6)/2}
    \nonumber \\
    &=
    \frac{45 C v^{-1+\alpha}}
    {2 \pi^2 g_{*s,\mathrm{b}} T_\mathrm{b}^3}
    \left( \frac{\pi^2 g_{*,\mathrm{osc}}}{10} \right)^{\frac{p+6}{2(p+4)}}
    \frac{\beta m_0 \phi_0^2 T_\mathrm{b}^{\frac{p+6}{2}}}
    { (m_0 M_\mathrm{Pl} \Lambda^{\frac{p}{2}})^{\frac{p+6}{p+4}} }
    \nonumber \\
    &=
    \frac{45 C v^{-1+\alpha} }
    {2 \pi^2 g_{*s,\mathrm{b}}}
    \left( \frac{\pi^2 g_{*,\mathrm{osc}}}{10} \right)^{\frac{p+6}{2(p+4)}}
    \frac{\beta m_0 \phi_0^2}
    {(m_\mathrm{b}^{p+6} M_\mathrm{Pl}^{p+6} T_\mathrm{b}^p)^{\frac{1}{p+4}}}
    \ .
    \label{eq: rho/s d2 min}
\end{align}
If $\phi = 0$ at the FOPT, the passage of the bubble wall does not change $\phi$ and $\dot{\phi}$.
In this case, the axion abundance becomes the same as in case (b2).
\begin{align}
    \frac{\rho_\phi^\mathrm{(d2\text{-}2)}}{s}
    &=
    \left. \frac{\rho_\phi^\mathrm{(b2)}}{s} \right|_{\sin \alpha_\mathrm{b} = 0}
    \ .
    \label{eq: rho/s d2 max}
\end{align}
Depending on the phase of the oscillation at the FOPT, the axion abundance will be between Eqs.~\eqref{eq: rho/s d2 min} and \eqref{eq: rho/s d2 max}.

We show the dependence of $\rho_\phi/s$ on $m_0$ in Fig.~\ref{fig: m0 vs rho}.
Here, we fixed $\beta/H_\mathrm{b}$ and $m_0/m_\mathrm{b}$ satisfying $\beta/(3H_\mathrm{b}) > m_0/m_\mathrm{b}$.
Then, case (b2) is realized, and if $\beta/(3H_\mathrm{b}) < m_0/m_\mathrm{b}$, case (d1) is realized instead of case (b2).
\begin{figure}[!t]
    \begin{center}  
        \includegraphics[width=0.45\textwidth]{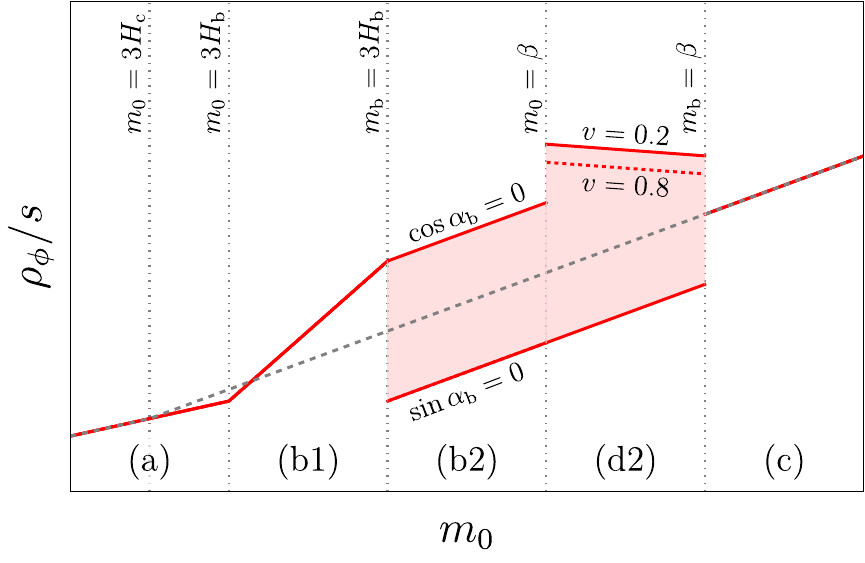}
        \end{center}
    \caption{%
        Axion abundance $\rho_\phi/s$.
        The red and gray-dashed lines represent the results for the first- and second-order phase transition, respectively. In the second-order phase transition case, $H_c$ is the Hubble parameter at $T=\Lambda$.
    }
    \label{fig: m0 vs rho} 
\end{figure}

Let us also estimate the axion-photon coupling $g_{\phi \gamma \gamma}$ to explain all dark matter.
If $\rho_\phi/s \simeq 0.44$\,eV, the axion accounts for all dark matter.
Once we fix {all} the parameters other than $\phi_0$, we obtain $\phi_0$ for all dark matter.
To relate the initial field amplitude to the axion-photon coupling, we assume $g_{\phi \gamma \gamma} = \alpha/(2 \pi f_\phi)$ and $\phi_0 = f_\phi$.
Then, we can obtain $g_{\phi \gamma \gamma}$ for axion dark matter by fixing the parameters of the axion mass and FOPT.

First, let us consider the case where $g_{\phi \gamma \gamma}$ takes the largest value. This corresponds to taking the lowest possible temperature at which the bubbles nucleate.
From the Lyman-$\alpha$ data, the redshift of the dark matter formation is constrained as $z \gtrsim z_\mathrm{min} \sim 10^6$~\cite{Sarkar:2014bca}.
Here, we require that the axion starts to oscillate and the axion mass becomes constant before $z = z_\mathrm{min}$.
Then, we obtain the minimum value of $f_\phi$ for axion dark matter, which leads to the maximum value of $g_{\phi \gamma \gamma}$.
We show the maximum $g_{\phi \gamma \gamma}$ for axion dark matter in the upper panel of Fig.~\ref{fig: m0 vs g}.
Here, we used $g_*$ and $g_{*s}$ given in Ref.~\cite{Husdal:2016haj}.
In case (a), the axion abundance is determined only by $m_0$ and $\phi_0$, and thus the line for case (a)  does not depend on $z_\mathrm{min}$.
In cases (b1), (b2), (c), (d1), and (d2-1), we obtained the upper bound by assuming $T_\mathrm{b} = T_\mathrm{min}$, where $T_\mathrm{min} \simeq 2.3 \times 10^2$\,eV is the cosmic temperature when $z =z_\mathrm{min}$.
For comparison, we also show the result for the second-order phase transition, where we assumed $\Lambda = T_\mathrm{min}$. 
Although $\rho_\phi^{\mathrm{(2nd)}}/s$ and $\rho_\phi^{\mathrm{(c)}}/s$ are given by the same formula, the upper bound for the second-order phase transition is slightly different from case (c), where we assume $T_\mathrm{b} = T_\mathrm{min}$.
Note that the constraint from the Lyman-$\alpha$ data can be more severe because the axions after the FOPT can be marginally relativistic and erase small-scale structures.
Thus, we can say that $T_\mathrm{b} = T_\mathrm{min}$ and $\Lambda = T_\mathrm{min}$ give optimistic bounds on $g_{\phi \gamma \gamma}$.
We also show the result using $10$\,MeV instead of $T_\mathrm{min}$ in the lower panel of Fig.~\ref{fig: m0 vs g}.
In both cases, we can see that the axion produced by the bubble misalignment mechanism can explain all dark matter in a region much larger than the conventional scenario represented by the lines of case (a) and `second'.
\begin{figure}[!t]
    \begin{center}  
        \includegraphics[width=0.45\textwidth]{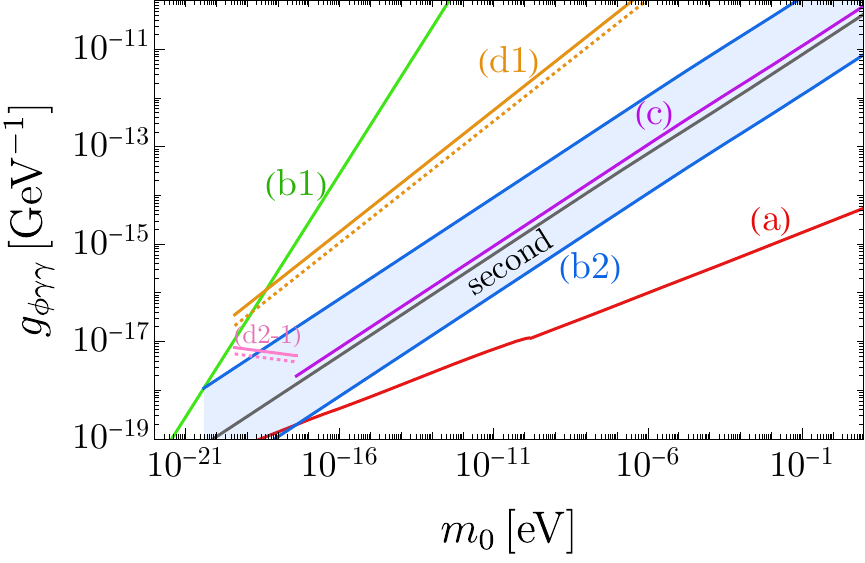}
        \\ \vspace{3mm}
        \includegraphics[width=0.45\textwidth]{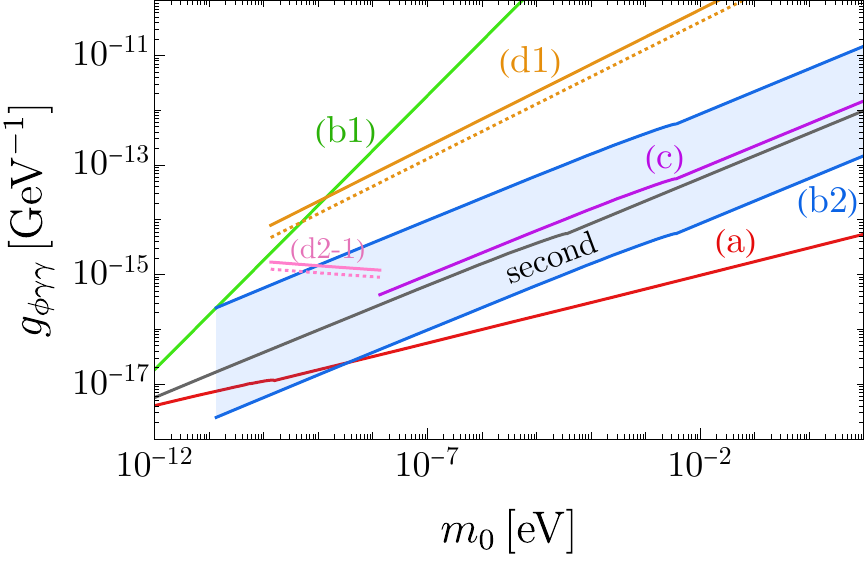}
        \end{center}
    \caption{%
        Upper bound on $g_{\phi \gamma \gamma} = \alpha/(2 \pi f_\phi)$ for axion dark matter in each scenario.
        We set $\phi_0 = f_\phi$ and $p = 8$.
        The red, green, blue, purple, orange, and pink lines represent cases (a), (b1), (b2), (c), (d1), and (d2-1), respectively.
        We use $T_\mathrm{b} = T_\mathrm{min} \simeq 2.3 \times 10^2$\,eV in the upper panel and $T_\mathrm{b} = 10$\,MeV in the lower panel to derive the upper bound for cases (b1), (b2), (c), (d1), and (d2-1).
        We also set $m_0/m_\mathrm{b} = 100$ for cases (b2), (c), and (d2-1), and $\beta/(3H_\mathrm{b}) = 10^3$ for cases (d1) and (d2-1).
        Cases (d1) and (d2-1) are shown in the solid and dotted lines for $v = 0.2$ and $v=0.8$, respectively.
        The blue-shaded region reflects the $\alpha_\mathrm{b}$-dependence of $\rho_\phi$ in case (b2).
        For comparison, we also show the result for the second-order phase transition assuming $\Lambda = T_\mathrm{min}$ (upper panel) or $\Lambda = 10$\,MeV (lower panel) with the gray line.
        Cases (b2), (c), and (d2) are shown only in the region where each case can be realized for $T_\mathrm{b} = T_\mathrm{min}$ or $10$\,MeV, $m_0/m_\mathrm{b} = 100$, and $\beta/(3H_\mathrm{b}) = 10^3$.
    }
    \label{fig: m0 vs g} 
\end{figure}

\section{Summary and discussions}
\label{sec: summary}
In this paper, we have studied the axion dynamics during the FOPT in which the axion mass changes discontinuously. In particular, we have taken into account the bubble wall dynamics for the first time and found that the axions can be expelled from the inside of the expanding bubbles when the axion mass inside the bubble is larger than $\beta$. This leads to a phenomenon analogous to Fermi acceleration, where the axion waves are repeatedly accelerated in collisions with the bubble walls until they gain sufficient energy to penetrate the wall.
Furthermore, we have found that the axion abundance is further increased by an order of magnitude when the axions enter the bubbles after repeated acceleration.
We have estimated the axion abundance, which turns out to be significantly larger than in the case where the axion mass is constant.

We have seen that the bubble dynamics induces a spatial inhomogeneity in the axion field. Whether such spatial inhomogeneity remains afterwards depends on the axion momentum and the size of the inhomogeneity.
The momentum distribution of the axions after the FOPT is marginally nonrelativistic.
Thus, for $\beta \gg H_{\rm b}$, the spatial inhomogeneities of axions generated during the FOPT are likely to be erased by the free streaming afterward. Depending on the detailed process during the FOPT, however, it might be possible to have relatively large inhomogeneities. For instance, if $\beta \sim H_{\rm b}$ and the bubble wall velocity is low, the resultant axions are nonrelativistic, and we expect that large inhomogeneities will persist on relatively large scales within the Hubble horizon, leading to the formation of the axion minicluster.
We note that the possibility of forming the axion minicluster in the FOPT was previously discussed in Ref.~\cite{Hardy:2016mns}, where it was studied that axions start to oscillate inside bubbles that arise at different times when the FOPT lasts the Hubble time or longer. This corresponds to case (b1), where the cosmic expansion and the (relatively long) duration of FOPT are included.

In our arguments, we have neglected the back reaction of the Fermi acceleration on the bubble wall dynamics. The axion waves may acquire large pressure through the scatterings, which slows down the expansion of the bubbles. Once the bubbles stop expanding, the Fermi acceleration will no longer occur, and the axion wave pressure and the bubble wall will be in equilibrium for a while. Eventually, the axion momentum will be redshifted by the cosmic expansion, and the bubbles will percolate. In this case, we expect a large spatial inhomogeneity of the axion dark matter to remain, which could become the axion minicluster. A detailed analysis is left for future studies.

Another immediate consequence of the nonzero modes excited during FOPT could be to hinder the formation of oscillons/I-balls~\cite{Bogolyubsky:1976nx,Segur:1987mg,Gleiser:1993pt,Copeland:1995fq,Gleiser:1999tj,Honda:2001xg,Kasuya:2002zs}, which are expected to occur efficiently if the onset of spatially homogeneous oscillations is delayed~\cite{Jeong:2022kdr,Nakagawa:2022wwm}. This is because the formation of these non-topological solitons requires coherence over a region much larger than the inverse of the axion mass.
On the other hand, the production of vector bosons coupled to axions may still proceed even in the presence of spatial inhomogeneities. Then, one may be able to produce the right amount of dark photon dark matter without introducing a large coupling, similar to Ref.~\cite{Kitajima:2023pby} (see also Refs.~\cite{Agrawal:2018vin, Co:2018lka,Bastero-Gil:2018uel}). One can also apply the axion dynamics during FOPTs to explain the recent hint of isotropic cosmic birefringence~\cite{Minami:2020odp,Diego-Palazuelos:2022dsq,Eskilt:2022wav,Eskilt:2022cff}.  For example, if the solar system is in a bubble formed after recombination and the axion waves have not yet transmitted much into the bubble,
we will observe isotropic cosmic birefringence. Also, if the bubbles nucleate during the recombination, it would lead to characteristic anisotropic cosmic birefringence. It will be interesting to explore the viable parameter range and estimate the power spectrum of the anisotropic cosmic birefringence.

While we have focused on the axion abundance, the dark glueballs could also contribute to dark matter. 
In case (b), in particular, the dark glueball abundance is considered to be comparable to the axion abundance, if $\phi_0$ is of order the decay constant.
As discussed in Ref.~\cite{Nakagawa:2022wwm}, the dark glueball abundance might be reduced by the coupling to the axion. One can also introduce dark quarks with dark photon couplings so that
the dark pion can decay or annihilate into the dark photons to cool the dark sector and reduce the glueball abundance.

So far we have considered the dynamics of the axion, whose mass changes significantly during the FOPT, but the same argument can be straightforwardly applied to any field whose mass changes during the FOPT or similar transitions induced by other fields.
In particular, the reflection and transmission processes, as well as the Fermi acceleration, also apply to fermions. Thus, various phenomena found in this paper may be quite general in the dynamics during phase transitions, and it could have interesting applications such as the production of warm dark matter, boson star formation, cosmic birefringence, baryogenesis, etc.

Let us comment on an implicit assumption made in this paper relevant to large $N$ Yang-Mills theory and the possible extension beyond the assumption.
In Yang-Mills theory coupled to the axion, the potential of the axion can be either single-valued or multi-valued~\cite{Witten:1980sp}.
So far we have implicitly assumed the former possibility. In particular, if we consider $\SU(N)$ Yang-Mills theory with large $N$, we may have the latter case, because in chiral perturbation theory, one expects a $(\log[\det U])^2$ term, which is multi-valued, for the $\eta'$ mass in the large $N$ limit, where $U$ is the matrix relevant for the Nambu-Goldstone boson.  
Let us comment on the cosmology with the latter possibility, which will turn out to be interesting. 
By integrating out the mesons, the axion potential is of the form
$V_{j}(a/(Nf_a))=V_0(a/(Nf_a)+2\pi j/N)$, where $j$ is the label relevant for different branches, with the combination $a/(Nf_a)$ appearing in the large $N$ limit. 
$V_{0}$ represents the potential form for each branch. 
Thanks to the multi-valued potential, the symmetry $a\to a+2\pi f_a$ is preserved, although it is violated for each potential.

Recovering the heavy meson potential, we find in the full potential that different branches are separated by potential barriers. Since $V_j$ has the same shape and minimum value for different $j$, this implies that the setup contains axion domain wall configurations. 
Thus, in cosmology, the domain walls may be produced and collapse due to the potential or population bias.  This contribution could affect the final abundance of axion dark matter, which will be an interesting topic for future studies in this direction. 

So far, we have had analytical discussions on the case that the wall width in the wall rest frame is much smaller than the axion wavelength. In the opposite limit, which, although is a slightly unnatural setup for a light axion dark matter, analytical estimations are also possible following Refs.~\cite{Azatov:2020ufh, Azatov:2021irb}. Since the wavelength is much shorter than the wall width, the shape of the wall is important for discussing the transmission rate or reflection rate. In this case, we can use the WKB approximation. Axion could be considered as a particle.
Still, the energy conservation is valid across the wall at the wall rest frame. The transmission/reflection probability can be estimated by quantum mechanics. For simplicity, when the momentum is larger than the axion mass scales, the amplitude can be approximated as ${\cal M}\sim \int \mathrm{d} z' \frac{1}{2 w'} \exp(i\Delta k' z' )m^2(z')$, with $\Delta k'$ being the difference of the momenta of incoming and outgoing waves in the rest frame, which can be obtained from the energy and mass outside the wall. $m^2(z')$ incorpolates the wall profile.
This is in sharp contrast to the transmission/reflection probabilities studied in the main text, where the transmission/reflection probabilities are determined by the boundary conditions of the axion wave, which are independent of the wall profile.

The more precise estimation of the axion abundance, the subsequent evolution of axions with spatial inhomogeneities, and the associated gravitational waves may require dedicated numerical lattice simulations taking into account the realistic bubble nucleation, which warrants further investigation in the future. It is also important to acknowledge the limitations of our analysis. Our study assumes a relatively fast FOPT, which may not be directly applicable to scenarios with very strong FOPTs, which could significantly affect the cosmological expansion. 
In addition, our analytical discussion assumes hierarchies among the time scales of axion oscillation, cosmic expansion, and FOPT.
If some of these time scales are comparable, our estimate of the axion abundance may not be applicable.
Moreover, we have simplified the complex dynamics of FOPTs by approximating the process with a single transition temperature, $T_\mathrm{b}$, neglecting, for example, the difference between the nucleation time and the percolation time.
Finally, the analysis of axion dynamics in the context of FOPTs neglected the effects of cosmic expansion, which could have a non-negligible impact on the results. These limitations highlight the need for more comprehensive models and simulations to fully understand the intricate dynamics of axions in the early universe, especially in the context of FOPTs.
 
\section*{Acknowledgments}
FT thanks Naoya Kitajima for useful discussions on vector boson production.
We are grateful to Giulio Barni for introducing our attention to his work~\cite{Azatov:2023xem}.
This work is supported by JSPS Core-to-Core Program (grant number: JPJSCCA20200002) (F.T.), JSPS KAKENHI Grant Numbers 20H01894 (F.T.), 20H05851 (F.T. and W.Y.), 21K20364 (W.Y.), 22K14029 (W.Y.), 22H01215 (W.Y.), 23KJ0088 (K.M.), Graduate Program on Physics for the Universe (J.L.), and Watanuki International Scholarship Foundation (J.L.).
This article is based upon work from COST Action COSMIC WISPers CA21106, supported by COST (European Cooperation in Science and Technology).

\bibliographystyle{apsrev4-1}
\bibliography{Ref}

\begin{thebibliography}{62}%
\makeatletter
\providecommand \@ifxundefined [1]{%
 \@ifx{#1\undefined}
}%
\providecommand \@ifnum [1]{%
 \ifnum #1\expandafter \@firstoftwo
 \else \expandafter \@secondoftwo
 \fi
}%
\providecommand \@ifx [1]{%
 \ifx #1\expandafter \@firstoftwo
 \else \expandafter \@secondoftwo
 \fi
}%
\providecommand \natexlab [1]{#1}%
\providecommand \enquote  [1]{``#1''}%
\providecommand \bibnamefont  [1]{#1}%
\providecommand \bibfnamefont [1]{#1}%
\providecommand \citenamefont [1]{#1}%
\providecommand \href@noop [0]{\@secondoftwo}%
\providecommand \href [0]{\begingroup \@sanitize@url \@href}%
\providecommand \@href[1]{\@@startlink{#1}\@@href}%
\providecommand \@@href[1]{\endgroup#1\@@endlink}%
\providecommand \@sanitize@url [0]{\catcode `\\12\catcode `\$12\catcode
  `\&12\catcode `\#12\catcode `\^12\catcode `\_12\catcode `\%12\relax}%
\providecommand \@@startlink[1]{}%
\providecommand \@@endlink[0]{}%
\providecommand \url  [0]{\begingroup\@sanitize@url \@url }%
\providecommand \@url [1]{\endgroup\@href {#1}{\urlprefix }}%
\providecommand \urlprefix  [0]{URL }%
\providecommand \Eprint [0]{\href }%
\providecommand \doibase [0]{http://dx.doi.org/}%
\providecommand \selectlanguage [0]{\@gobble}%
\providecommand \bibinfo  [0]{\@secondoftwo}%
\providecommand \bibfield  [0]{\@secondoftwo}%
\providecommand \translation [1]{[#1]}%
\providecommand \BibitemOpen [0]{}%
\providecommand \bibitemStop [0]{}%
\providecommand \bibitemNoStop [0]{.\EOS\space}%
\providecommand \EOS [0]{\spacefactor3000\relax}%
\providecommand \BibitemShut  [1]{\csname bibitem#1\endcsname}%
\let\auto@bib@innerbib\@empty
\bibitem [{\citenamefont {Peccei}\ and\ \citenamefont
  {Quinn}(1977{\natexlab{a}})}]{Peccei:1977hh}%
  \BibitemOpen
  \bibfield  {author} {\bibinfo {author} {\bibfnamefont {R.~D.}\ \bibnamefont
  {Peccei}}\ and\ \bibinfo {author} {\bibfnamefont {H.~R.}\ \bibnamefont
  {Quinn}},\ }\href {\doibase 10.1103/PhysRevLett.38.1440} {\bibfield
  {journal} {\bibinfo  {journal} {Phys. Rev. Lett.}\ }\textbf {\bibinfo
  {volume} {38}},\ \bibinfo {pages} {1440} (\bibinfo {year}
  {1977}{\natexlab{a}})}\BibitemShut {NoStop}%
\bibitem [{\citenamefont {Peccei}\ and\ \citenamefont
  {Quinn}(1977{\natexlab{b}})}]{Peccei:1977ur}%
  \BibitemOpen
  \bibfield  {author} {\bibinfo {author} {\bibfnamefont {R.~D.}\ \bibnamefont
  {Peccei}}\ and\ \bibinfo {author} {\bibfnamefont {H.~R.}\ \bibnamefont
  {Quinn}},\ }\href {\doibase 10.1103/PhysRevD.16.1791} {\bibfield  {journal}
  {\bibinfo  {journal} {Phys. Rev. D}\ }\textbf {\bibinfo {volume} {16}},\
  \bibinfo {pages} {1791} (\bibinfo {year} {1977}{\natexlab{b}})}\BibitemShut
  {NoStop}%
\bibitem [{\citenamefont {Weinberg}(1978)}]{Weinberg:1977ma}%
  \BibitemOpen
  \bibfield  {author} {\bibinfo {author} {\bibfnamefont {S.}~\bibnamefont
  {Weinberg}},\ }\href {\doibase 10.1103/PhysRevLett.40.223} {\bibfield
  {journal} {\bibinfo  {journal} {Phys. Rev. Lett.}\ }\textbf {\bibinfo
  {volume} {40}},\ \bibinfo {pages} {223} (\bibinfo {year} {1978})}\BibitemShut
  {NoStop}%
\bibitem [{\citenamefont {Wilczek}(1978)}]{Wilczek:1977pj}%
  \BibitemOpen
  \bibfield  {author} {\bibinfo {author} {\bibfnamefont {F.}~\bibnamefont
  {Wilczek}},\ }\href {\doibase 10.1103/PhysRevLett.40.279} {\bibfield
  {journal} {\bibinfo  {journal} {Phys. Rev. Lett.}\ }\textbf {\bibinfo
  {volume} {40}},\ \bibinfo {pages} {279} (\bibinfo {year} {1978})}\BibitemShut
  {NoStop}%
\bibitem [{\citenamefont {Preskill}\ \emph {et~al.}(1983)\citenamefont
  {Preskill}, \citenamefont {Wise},\ and\ \citenamefont
  {Wilczek}}]{Preskill:1982cy}%
  \BibitemOpen
  \bibfield  {author} {\bibinfo {author} {\bibfnamefont {J.}~\bibnamefont
  {Preskill}}, \bibinfo {author} {\bibfnamefont {M.~B.}\ \bibnamefont {Wise}},
  \ and\ \bibinfo {author} {\bibfnamefont {F.}~\bibnamefont {Wilczek}},\ }\href
  {\doibase 10.1016/0370-2693(83)90637-8} {\bibfield  {journal} {\bibinfo
  {journal} {Phys. Lett. B}\ }\textbf {\bibinfo {volume} {120}},\ \bibinfo
  {pages} {127} (\bibinfo {year} {1983})}\BibitemShut {NoStop}%
\bibitem [{\citenamefont {Abbott}\ and\ \citenamefont
  {Sikivie}(1983)}]{Abbott:1982af}%
  \BibitemOpen
  \bibfield  {author} {\bibinfo {author} {\bibfnamefont {L.~F.}\ \bibnamefont
  {Abbott}}\ and\ \bibinfo {author} {\bibfnamefont {P.}~\bibnamefont
  {Sikivie}},\ }\href {\doibase 10.1016/0370-2693(83)90638-X} {\bibfield
  {journal} {\bibinfo  {journal} {Phys. Lett. B}\ }\textbf {\bibinfo {volume}
  {120}},\ \bibinfo {pages} {133} (\bibinfo {year} {1983})}\BibitemShut
  {NoStop}%
\bibitem [{\citenamefont {Dine}\ and\ \citenamefont
  {Fischler}(1983)}]{Dine:1982ah}%
  \BibitemOpen
  \bibfield  {author} {\bibinfo {author} {\bibfnamefont {M.}~\bibnamefont
  {Dine}}\ and\ \bibinfo {author} {\bibfnamefont {W.}~\bibnamefont
  {Fischler}},\ }\href {\doibase 10.1016/0370-2693(83)90639-1} {\bibfield
  {journal} {\bibinfo  {journal} {Phys. Lett. B}\ }\textbf {\bibinfo {volume}
  {120}},\ \bibinfo {pages} {137} (\bibinfo {year} {1983})}\BibitemShut
  {NoStop}%
\bibitem [{\citenamefont {Higaki}\ \emph {et~al.}(2016)\citenamefont {Higaki},
  \citenamefont {Jeong}, \citenamefont {Kitajima},\ and\ \citenamefont
  {Takahashi}}]{Higaki:2016yqk}%
  \BibitemOpen
  \bibfield  {author} {\bibinfo {author} {\bibfnamefont {T.}~\bibnamefont
  {Higaki}}, \bibinfo {author} {\bibfnamefont {K.~S.}\ \bibnamefont {Jeong}},
  \bibinfo {author} {\bibfnamefont {N.}~\bibnamefont {Kitajima}}, \ and\
  \bibinfo {author} {\bibfnamefont {F.}~\bibnamefont {Takahashi}},\ }\href
  {\doibase 10.1007/JHEP06(2016)150} {\bibfield  {journal} {\bibinfo  {journal}
  {JHEP}\ }\textbf {\bibinfo {volume} {06}},\ \bibinfo {pages} {150} (\bibinfo
  {year} {2016})},\ \Eprint {http://arxiv.org/abs/1603.02090} {arXiv:1603.02090
  [hep-ph]} \BibitemShut {NoStop}%
\bibitem [{\citenamefont {Di~Luzio}\ \emph {et~al.}(2021)\citenamefont
  {Di~Luzio}, \citenamefont {Gavela}, \citenamefont {Quilez},\ and\
  \citenamefont {Ringwald}}]{DiLuzio:2021gos}%
  \BibitemOpen
  \bibfield  {author} {\bibinfo {author} {\bibfnamefont {L.}~\bibnamefont
  {Di~Luzio}}, \bibinfo {author} {\bibfnamefont {B.}~\bibnamefont {Gavela}},
  \bibinfo {author} {\bibfnamefont {P.}~\bibnamefont {Quilez}}, \ and\ \bibinfo
  {author} {\bibfnamefont {A.}~\bibnamefont {Ringwald}},\ }\href {\doibase
  10.1088/1475-7516/2021/10/001} {\bibfield  {journal} {\bibinfo  {journal}
  {JCAP}\ }\textbf {\bibinfo {volume} {10}},\ \bibinfo {pages} {001} (\bibinfo
  {year} {2021})},\ \Eprint {http://arxiv.org/abs/2102.01082} {arXiv:2102.01082
  [hep-ph]} \BibitemShut {NoStop}%
\bibitem [{\citenamefont {Jeong}\ \emph {et~al.}(2022)\citenamefont {Jeong},
  \citenamefont {Matsukawa}, \citenamefont {Nakagawa},\ and\ \citenamefont
  {Takahashi}}]{Jeong:2022kdr}%
  \BibitemOpen
  \bibfield  {author} {\bibinfo {author} {\bibfnamefont {K.~S.}\ \bibnamefont
  {Jeong}}, \bibinfo {author} {\bibfnamefont {K.}~\bibnamefont {Matsukawa}},
  \bibinfo {author} {\bibfnamefont {S.}~\bibnamefont {Nakagawa}}, \ and\
  \bibinfo {author} {\bibfnamefont {F.}~\bibnamefont {Takahashi}},\ }\href
  {\doibase 10.1088/1475-7516/2022/03/026} {\bibfield  {journal} {\bibinfo
  {journal} {JCAP}\ }\textbf {\bibinfo {volume} {03}},\ \bibinfo {pages} {026}
  (\bibinfo {year} {2022})},\ \Eprint {http://arxiv.org/abs/2201.00681}
  {arXiv:2201.00681 [hep-ph]} \BibitemShut {NoStop}%
\bibitem [{\citenamefont {Daido}\ \emph {et~al.}(2017)\citenamefont {Daido},
  \citenamefont {Takahashi},\ and\ \citenamefont {Yin}}]{Daido:2017wwb}%
  \BibitemOpen
  \bibfield  {author} {\bibinfo {author} {\bibfnamefont {R.}~\bibnamefont
  {Daido}}, \bibinfo {author} {\bibfnamefont {F.}~\bibnamefont {Takahashi}}, \
  and\ \bibinfo {author} {\bibfnamefont {W.}~\bibnamefont {Yin}},\ }\href
  {\doibase 10.1088/1475-7516/2017/05/044} {\bibfield  {journal} {\bibinfo
  {journal} {JCAP}\ }\textbf {\bibinfo {volume} {05}},\ \bibinfo {pages} {044}
  (\bibinfo {year} {2017})},\ \Eprint {http://arxiv.org/abs/1702.03284}
  {arXiv:1702.03284 [hep-ph]} \BibitemShut {NoStop}%
\bibitem [{\citenamefont {Takahashi}\ and\ \citenamefont
  {Yin}(2019)}]{Takahashi:2019pqf}%
  \BibitemOpen
  \bibfield  {author} {\bibinfo {author} {\bibfnamefont {F.}~\bibnamefont
  {Takahashi}}\ and\ \bibinfo {author} {\bibfnamefont {W.}~\bibnamefont
  {Yin}},\ }\href {\doibase 10.1007/JHEP10(2019)120} {\bibfield  {journal}
  {\bibinfo  {journal} {JHEP}\ }\textbf {\bibinfo {volume} {10}},\ \bibinfo
  {pages} {120} (\bibinfo {year} {2019})},\ \Eprint
  {http://arxiv.org/abs/1908.06071} {arXiv:1908.06071 [hep-ph]} \BibitemShut
  {NoStop}%
\bibitem [{\citenamefont {Nakagawa}\ \emph {et~al.}(2020)\citenamefont
  {Nakagawa}, \citenamefont {Takahashi},\ and\ \citenamefont
  {Yin}}]{Nakagawa:2020eeg}%
  \BibitemOpen
  \bibfield  {author} {\bibinfo {author} {\bibfnamefont {S.}~\bibnamefont
  {Nakagawa}}, \bibinfo {author} {\bibfnamefont {F.}~\bibnamefont {Takahashi}},
  \ and\ \bibinfo {author} {\bibfnamefont {W.}~\bibnamefont {Yin}},\ }\href
  {\doibase 10.1088/1475-7516/2020/05/004} {\bibfield  {journal} {\bibinfo
  {journal} {JCAP}\ }\textbf {\bibinfo {volume} {05}},\ \bibinfo {pages} {004}
  (\bibinfo {year} {2020})},\ \Eprint {http://arxiv.org/abs/2002.12195}
  {arXiv:2002.12195 [hep-ph]} \BibitemShut {NoStop}%
\bibitem [{\citenamefont {Narita}\ \emph {et~al.}(2023)\citenamefont {Narita},
  \citenamefont {Takahashi},\ and\ \citenamefont {Yin}}]{Narita:2023naj}%
  \BibitemOpen
  \bibfield  {author} {\bibinfo {author} {\bibfnamefont {Y.}~\bibnamefont
  {Narita}}, \bibinfo {author} {\bibfnamefont {F.}~\bibnamefont {Takahashi}}, \
  and\ \bibinfo {author} {\bibfnamefont {W.}~\bibnamefont {Yin}},\ }\href
  {\doibase 10.1088/1475-7516/2023/12/039} {\bibfield  {journal} {\bibinfo
  {journal} {JCAP}\ }\textbf {\bibinfo {volume} {12}},\ \bibinfo {pages} {039}
  (\bibinfo {year} {2023})},\ \Eprint {http://arxiv.org/abs/2308.12154}
  {arXiv:2308.12154 [hep-ph]} \BibitemShut {NoStop}%
\bibitem [{\citenamefont {Kitajima}\ and\ \citenamefont
  {Takahashi}(2015)}]{Kitajima:2014xla}%
  \BibitemOpen
  \bibfield  {author} {\bibinfo {author} {\bibfnamefont {N.}~\bibnamefont
  {Kitajima}}\ and\ \bibinfo {author} {\bibfnamefont {F.}~\bibnamefont
  {Takahashi}},\ }\href {\doibase 10.1088/1475-7516/2015/01/032} {\bibfield
  {journal} {\bibinfo  {journal} {JCAP}\ }\textbf {\bibinfo {volume} {01}},\
  \bibinfo {pages} {032} (\bibinfo {year} {2015})},\ \Eprint
  {http://arxiv.org/abs/1411.2011} {arXiv:1411.2011 [hep-ph]} \BibitemShut
  {NoStop}%
\bibitem [{\citenamefont {Daido}\ \emph {et~al.}(2015)\citenamefont {Daido},
  \citenamefont {Kitajima},\ and\ \citenamefont {Takahashi}}]{Daido:2015bva}%
  \BibitemOpen
  \bibfield  {author} {\bibinfo {author} {\bibfnamefont {R.}~\bibnamefont
  {Daido}}, \bibinfo {author} {\bibfnamefont {N.}~\bibnamefont {Kitajima}}, \
  and\ \bibinfo {author} {\bibfnamefont {F.}~\bibnamefont {Takahashi}},\ }\href
  {\doibase 10.1103/PhysRevD.92.063512} {\bibfield  {journal} {\bibinfo
  {journal} {Phys. Rev. D}\ }\textbf {\bibinfo {volume} {92}},\ \bibinfo
  {pages} {063512} (\bibinfo {year} {2015})},\ \Eprint
  {http://arxiv.org/abs/1505.07670} {arXiv:1505.07670 [hep-ph]} \BibitemShut
  {NoStop}%
\bibitem [{\citenamefont {Daido}\ \emph {et~al.}(2016)\citenamefont {Daido},
  \citenamefont {Kitajima},\ and\ \citenamefont {Takahashi}}]{Daido:2015cba}%
  \BibitemOpen
  \bibfield  {author} {\bibinfo {author} {\bibfnamefont {R.}~\bibnamefont
  {Daido}}, \bibinfo {author} {\bibfnamefont {N.}~\bibnamefont {Kitajima}}, \
  and\ \bibinfo {author} {\bibfnamefont {F.}~\bibnamefont {Takahashi}},\ }\href
  {\doibase 10.1103/PhysRevD.93.075027} {\bibfield  {journal} {\bibinfo
  {journal} {Phys. Rev. D}\ }\textbf {\bibinfo {volume} {93}},\ \bibinfo
  {pages} {075027} (\bibinfo {year} {2016})},\ \Eprint
  {http://arxiv.org/abs/1510.06675} {arXiv:1510.06675 [hep-ph]} \BibitemShut
  {NoStop}%
\bibitem [{\citenamefont {Ho}\ \emph {et~al.}(2018)\citenamefont {Ho},
  \citenamefont {Saikawa},\ and\ \citenamefont {Takahashi}}]{Ho:2018qur}%
  \BibitemOpen
  \bibfield  {author} {\bibinfo {author} {\bibfnamefont {S.-Y.}\ \bibnamefont
  {Ho}}, \bibinfo {author} {\bibfnamefont {K.}~\bibnamefont {Saikawa}}, \ and\
  \bibinfo {author} {\bibfnamefont {F.}~\bibnamefont {Takahashi}},\ }\href
  {\doibase 10.1088/1475-7516/2018/10/042} {\bibfield  {journal} {\bibinfo
  {journal} {JCAP}\ }\textbf {\bibinfo {volume} {10}},\ \bibinfo {pages} {042}
  (\bibinfo {year} {2018})},\ \Eprint {http://arxiv.org/abs/1806.09551}
  {arXiv:1806.09551 [hep-ph]} \BibitemShut {NoStop}%
\bibitem [{\citenamefont {Murai}\ \emph {et~al.}(2023)\citenamefont {Murai},
  \citenamefont {Takahashi},\ and\ \citenamefont {Yin}}]{Murai:2023xjn}%
  \BibitemOpen
  \bibfield  {author} {\bibinfo {author} {\bibfnamefont {K.}~\bibnamefont
  {Murai}}, \bibinfo {author} {\bibfnamefont {F.}~\bibnamefont {Takahashi}}, \
  and\ \bibinfo {author} {\bibfnamefont {W.}~\bibnamefont {Yin}},\ }\href
  {\doibase 10.1103/PhysRevD.108.036020} {\bibfield  {journal} {\bibinfo
  {journal} {Phys. Rev. D}\ }\textbf {\bibinfo {volume} {108}},\ \bibinfo
  {pages} {036020} (\bibinfo {year} {2023})},\ \Eprint
  {http://arxiv.org/abs/2305.18677} {arXiv:2305.18677 [hep-ph]} \BibitemShut
  {NoStop}%
\bibitem [{\citenamefont {Nakagawa}\ \emph {et~al.}(2023)\citenamefont
  {Nakagawa}, \citenamefont {Takahashi}, \citenamefont {Yamada},\ and\
  \citenamefont {Yin}}]{Nakagawa:2022wwm}%
  \BibitemOpen
  \bibfield  {author} {\bibinfo {author} {\bibfnamefont {S.}~\bibnamefont
  {Nakagawa}}, \bibinfo {author} {\bibfnamefont {F.}~\bibnamefont {Takahashi}},
  \bibinfo {author} {\bibfnamefont {M.}~\bibnamefont {Yamada}}, \ and\ \bibinfo
  {author} {\bibfnamefont {W.}~\bibnamefont {Yin}},\ }\href {\doibase
  10.1016/j.physletb.2023.137824} {\bibfield  {journal} {\bibinfo  {journal}
  {Phys. Lett. B}\ }\textbf {\bibinfo {volume} {839}},\ \bibinfo {pages}
  {137824} (\bibinfo {year} {2023})},\ \Eprint
  {http://arxiv.org/abs/2210.10022} {arXiv:2210.10022 [hep-ph]} \BibitemShut
  {NoStop}%
\bibitem [{\citenamefont {Cyncynates}\ and\ \citenamefont
  {Thompson}(2023)}]{Cyncynates:2023esj}%
  \BibitemOpen
  \bibfield  {author} {\bibinfo {author} {\bibfnamefont {D.}~\bibnamefont
  {Cyncynates}}\ and\ \bibinfo {author} {\bibfnamefont {J.~O.}\ \bibnamefont
  {Thompson}},\ }\href {\doibase 10.1103/PhysRevD.108.L091703} {\bibfield
  {journal} {\bibinfo  {journal} {Phys. Rev. D}\ }\textbf {\bibinfo {volume}
  {108}},\ \bibinfo {pages} {L091703} (\bibinfo {year} {2023})},\ \Eprint
  {http://arxiv.org/abs/2306.04678} {arXiv:2306.04678 [hep-ph]} \BibitemShut
  {NoStop}%
\bibitem [{\citenamefont {Co}\ \emph {et~al.}(2020)\citenamefont {Co},
  \citenamefont {Hall},\ and\ \citenamefont {Harigaya}}]{Co:2019jts}%
  \BibitemOpen
  \bibfield  {author} {\bibinfo {author} {\bibfnamefont {R.~T.}\ \bibnamefont
  {Co}}, \bibinfo {author} {\bibfnamefont {L.~J.}\ \bibnamefont {Hall}}, \ and\
  \bibinfo {author} {\bibfnamefont {K.}~\bibnamefont {Harigaya}},\ }\href
  {\doibase 10.1103/PhysRevLett.124.251802} {\bibfield  {journal} {\bibinfo
  {journal} {Phys. Rev. Lett.}\ }\textbf {\bibinfo {volume} {124}},\ \bibinfo
  {pages} {251802} (\bibinfo {year} {2020})},\ \Eprint
  {http://arxiv.org/abs/1910.14152} {arXiv:1910.14152 [hep-ph]} \BibitemShut
  {NoStop}%
\bibitem [{\citenamefont {Lucini}\ \emph {et~al.}(2004)\citenamefont {Lucini},
  \citenamefont {Teper},\ and\ \citenamefont {Wenger}}]{Lucini:2003zr}%
  \BibitemOpen
  \bibfield  {author} {\bibinfo {author} {\bibfnamefont {B.}~\bibnamefont
  {Lucini}}, \bibinfo {author} {\bibfnamefont {M.}~\bibnamefont {Teper}}, \
  and\ \bibinfo {author} {\bibfnamefont {U.}~\bibnamefont {Wenger}},\ }\href
  {\doibase 10.1088/1126-6708/2004/01/061} {\bibfield  {journal} {\bibinfo
  {journal} {JHEP}\ }\textbf {\bibinfo {volume} {01}},\ \bibinfo {pages} {061}
  (\bibinfo {year} {2004})},\ \Eprint {http://arxiv.org/abs/hep-lat/0307017}
  {arXiv:hep-lat/0307017} \BibitemShut {NoStop}%
\bibitem [{\citenamefont {Lucini}\ \emph {et~al.}(2005)\citenamefont {Lucini},
  \citenamefont {Teper},\ and\ \citenamefont {Wenger}}]{Lucini:2005vg}%
  \BibitemOpen
  \bibfield  {author} {\bibinfo {author} {\bibfnamefont {B.}~\bibnamefont
  {Lucini}}, \bibinfo {author} {\bibfnamefont {M.}~\bibnamefont {Teper}}, \
  and\ \bibinfo {author} {\bibfnamefont {U.}~\bibnamefont {Wenger}},\ }\href
  {\doibase 10.1088/1126-6708/2005/02/033} {\bibfield  {journal} {\bibinfo
  {journal} {JHEP}\ }\textbf {\bibinfo {volume} {02}},\ \bibinfo {pages} {033}
  (\bibinfo {year} {2005})},\ \Eprint {http://arxiv.org/abs/hep-lat/0502003}
  {arXiv:hep-lat/0502003} \BibitemShut {NoStop}%
\bibitem [{\citenamefont {Blandford}\ and\ \citenamefont
  {Ostriker}(1978)}]{Blandford:1978ky}%
  \BibitemOpen
  \bibfield  {author} {\bibinfo {author} {\bibfnamefont {R.~D.}\ \bibnamefont
  {Blandford}}\ and\ \bibinfo {author} {\bibfnamefont {J.~P.}\ \bibnamefont
  {Ostriker}},\ }\href {\doibase 10.1086/182658} {\bibfield  {journal}
  {\bibinfo  {journal} {Astrophys. J. Lett.}\ }\textbf {\bibinfo {volume}
  {221}},\ \bibinfo {pages} {L29} (\bibinfo {year} {1978})}\BibitemShut
  {NoStop}%
\bibitem [{\citenamefont {Bell}(1978)}]{Bell:1978clk}%
  \BibitemOpen
  \bibfield  {author} {\bibinfo {author} {\bibfnamefont {A.~R.}\ \bibnamefont
  {Bell}},\ }\href {\doibase 10.1093/mnras/182.2.147} {\bibfield  {journal}
  {\bibinfo  {journal} {Mon. Not. Roy. Astron. Soc.}\ }\textbf {\bibinfo
  {volume} {182}},\ \bibinfo {pages} {147} (\bibinfo {year}
  {1978})}\BibitemShut {NoStop}%
\bibitem [{\citenamefont {Drury}(1983)}]{Drury:1983zz}%
  \BibitemOpen
  \bibfield  {author} {\bibinfo {author} {\bibfnamefont {L.~O.}\ \bibnamefont
  {Drury}},\ }\href {\doibase 10.1088/0034-4885/46/8/002} {\bibfield  {journal}
  {\bibinfo  {journal} {Rept. Prog. Phys.}\ }\textbf {\bibinfo {volume} {46}},\
  \bibinfo {pages} {973} (\bibinfo {year} {1983})}\BibitemShut {NoStop}%
\bibitem [{\citenamefont {Blandford}\ and\ \citenamefont
  {Eichler}(1987)}]{Blandford:1987pw}%
  \BibitemOpen
  \bibfield  {author} {\bibinfo {author} {\bibfnamefont {R.}~\bibnamefont
  {Blandford}}\ and\ \bibinfo {author} {\bibfnamefont {D.}~\bibnamefont
  {Eichler}},\ }\href {\doibase 10.1016/0370-1573(87)90134-7} {\bibfield
  {journal} {\bibinfo  {journal} {Phys. Rept.}\ }\textbf {\bibinfo {volume}
  {154}},\ \bibinfo {pages} {1} (\bibinfo {year} {1987})}\BibitemShut {NoStop}%
\bibitem [{\citenamefont {Baker}\ \emph {et~al.}(2020)\citenamefont {Baker},
  \citenamefont {Kopp},\ and\ \citenamefont {Long}}]{Baker:2019ndr}%
  \BibitemOpen
  \bibfield  {author} {\bibinfo {author} {\bibfnamefont {M.~J.}\ \bibnamefont
  {Baker}}, \bibinfo {author} {\bibfnamefont {J.}~\bibnamefont {Kopp}}, \ and\
  \bibinfo {author} {\bibfnamefont {A.~J.}\ \bibnamefont {Long}},\ }\href
  {\doibase 10.1103/PhysRevLett.125.151102} {\bibfield  {journal} {\bibinfo
  {journal} {Phys. Rev. Lett.}\ }\textbf {\bibinfo {volume} {125}},\ \bibinfo
  {pages} {151102} (\bibinfo {year} {2020})},\ \Eprint
  {http://arxiv.org/abs/1912.02830} {arXiv:1912.02830 [hep-ph]} \BibitemShut
  {NoStop}%
\bibitem [{\citenamefont {Chway}\ \emph {et~al.}(2020)\citenamefont {Chway},
  \citenamefont {Jung},\ and\ \citenamefont {Shin}}]{Chway:2019kft}%
  \BibitemOpen
  \bibfield  {author} {\bibinfo {author} {\bibfnamefont {D.}~\bibnamefont
  {Chway}}, \bibinfo {author} {\bibfnamefont {T.~H.}\ \bibnamefont {Jung}}, \
  and\ \bibinfo {author} {\bibfnamefont {C.~S.}\ \bibnamefont {Shin}},\ }\href
  {\doibase 10.1103/PhysRevD.101.095019} {\bibfield  {journal} {\bibinfo
  {journal} {Phys. Rev. D}\ }\textbf {\bibinfo {volume} {101}},\ \bibinfo
  {pages} {095019} (\bibinfo {year} {2020})},\ \Eprint
  {http://arxiv.org/abs/1912.04238} {arXiv:1912.04238 [hep-ph]} \BibitemShut
  {NoStop}%
\bibitem [{\citenamefont {Azatov}\ \emph
  {et~al.}(2021{\natexlab{a}})\citenamefont {Azatov}, \citenamefont
  {Vanvlasselaer},\ and\ \citenamefont {Yin}}]{Azatov:2021ifm}%
  \BibitemOpen
  \bibfield  {author} {\bibinfo {author} {\bibfnamefont {A.}~\bibnamefont
  {Azatov}}, \bibinfo {author} {\bibfnamefont {M.}~\bibnamefont
  {Vanvlasselaer}}, \ and\ \bibinfo {author} {\bibfnamefont {W.}~\bibnamefont
  {Yin}},\ }\href {\doibase 10.1007/JHEP03(2021)288} {\bibfield  {journal}
  {\bibinfo  {journal} {JHEP}\ }\textbf {\bibinfo {volume} {03}},\ \bibinfo
  {pages} {288} (\bibinfo {year} {2021}{\natexlab{a}})},\ \Eprint
  {http://arxiv.org/abs/2101.05721} {arXiv:2101.05721 [hep-ph]} \BibitemShut
  {NoStop}%
\bibitem [{\citenamefont {Baldes}\ \emph {et~al.}(2023)\citenamefont {Baldes},
  \citenamefont {Gouttenoire},\ and\ \citenamefont {Sala}}]{Baldes:2022oev}%
  \BibitemOpen
  \bibfield  {author} {\bibinfo {author} {\bibfnamefont {I.}~\bibnamefont
  {Baldes}}, \bibinfo {author} {\bibfnamefont {Y.}~\bibnamefont {Gouttenoire}},
  \ and\ \bibinfo {author} {\bibfnamefont {F.}~\bibnamefont {Sala}},\ }\href
  {\doibase 10.21468/SciPostPhys.14.3.033} {\bibfield  {journal} {\bibinfo
  {journal} {SciPost Phys.}\ }\textbf {\bibinfo {volume} {14}},\ \bibinfo
  {pages} {033} (\bibinfo {year} {2023})},\ \Eprint
  {http://arxiv.org/abs/2207.05096} {arXiv:2207.05096 [hep-ph]} \BibitemShut
  {NoStop}%
\bibitem [{\citenamefont {Azatov}\ \emph {et~al.}(2022)\citenamefont {Azatov},
  \citenamefont {Barni}, \citenamefont {Chakraborty}, \citenamefont
  {Vanvlasselaer},\ and\ \citenamefont {Yin}}]{Azatov:2022tii}%
  \BibitemOpen
  \bibfield  {author} {\bibinfo {author} {\bibfnamefont {A.}~\bibnamefont
  {Azatov}}, \bibinfo {author} {\bibfnamefont {G.}~\bibnamefont {Barni}},
  \bibinfo {author} {\bibfnamefont {S.}~\bibnamefont {Chakraborty}}, \bibinfo
  {author} {\bibfnamefont {M.}~\bibnamefont {Vanvlasselaer}}, \ and\ \bibinfo
  {author} {\bibfnamefont {W.}~\bibnamefont {Yin}},\ }\href {\doibase
  10.1007/JHEP10(2022)017} {\bibfield  {journal} {\bibinfo  {journal} {JHEP}\
  }\textbf {\bibinfo {volume} {10}},\ \bibinfo {pages} {017} (\bibinfo {year}
  {2022})},\ \Eprint {http://arxiv.org/abs/2207.02230} {arXiv:2207.02230
  [hep-ph]} \BibitemShut {NoStop}%
\bibitem [{\citenamefont {Falkowski}\ and\ \citenamefont
  {No}(2013)}]{Falkowski:2012fb}%
  \BibitemOpen
  \bibfield  {author} {\bibinfo {author} {\bibfnamefont {A.}~\bibnamefont
  {Falkowski}}\ and\ \bibinfo {author} {\bibfnamefont {J.~M.}\ \bibnamefont
  {No}},\ }\href {\doibase 10.1007/JHEP02(2013)034} {\bibfield  {journal}
  {\bibinfo  {journal} {JHEP}\ }\textbf {\bibinfo {volume} {02}},\ \bibinfo
  {pages} {034} (\bibinfo {year} {2013})},\ \Eprint
  {http://arxiv.org/abs/1211.5615} {arXiv:1211.5615 [hep-ph]} \BibitemShut
  {NoStop}%
\bibitem [{\citenamefont {Bodeker}\ and\ \citenamefont
  {Moore}(2009)}]{Bodeker:2009qy}%
  \BibitemOpen
  \bibfield  {author} {\bibinfo {author} {\bibfnamefont {D.}~\bibnamefont
  {Bodeker}}\ and\ \bibinfo {author} {\bibfnamefont {G.~D.}\ \bibnamefont
  {Moore}},\ }\href {\doibase 10.1088/1475-7516/2009/05/009} {\bibfield
  {journal} {\bibinfo  {journal} {JCAP}\ }\textbf {\bibinfo {volume} {05}},\
  \bibinfo {pages} {009} (\bibinfo {year} {2009})},\ \Eprint
  {http://arxiv.org/abs/0903.4099} {arXiv:0903.4099 [hep-ph]} \BibitemShut
  {NoStop}%
\bibitem [{\citenamefont {Bodeker}\ and\ \citenamefont
  {Moore}(2017)}]{Bodeker:2017cim}%
  \BibitemOpen
  \bibfield  {author} {\bibinfo {author} {\bibfnamefont {D.}~\bibnamefont
  {Bodeker}}\ and\ \bibinfo {author} {\bibfnamefont {G.~D.}\ \bibnamefont
  {Moore}},\ }\href {\doibase 10.1088/1475-7516/2017/05/025} {\bibfield
  {journal} {\bibinfo  {journal} {JCAP}\ }\textbf {\bibinfo {volume} {05}},\
  \bibinfo {pages} {025} (\bibinfo {year} {2017})},\ \Eprint
  {http://arxiv.org/abs/1703.08215} {arXiv:1703.08215 [hep-ph]} \BibitemShut
  {NoStop}%
\bibitem [{\citenamefont {H\"oche}\ \emph {et~al.}(2021)\citenamefont
  {H\"oche}, \citenamefont {Kozaczuk}, \citenamefont {Long}, \citenamefont
  {Turner},\ and\ \citenamefont {Wang}}]{Hoche:2020ysm}%
  \BibitemOpen
  \bibfield  {author} {\bibinfo {author} {\bibfnamefont {S.}~\bibnamefont
  {H\"oche}}, \bibinfo {author} {\bibfnamefont {J.}~\bibnamefont {Kozaczuk}},
  \bibinfo {author} {\bibfnamefont {A.~J.}\ \bibnamefont {Long}}, \bibinfo
  {author} {\bibfnamefont {J.}~\bibnamefont {Turner}}, \ and\ \bibinfo {author}
  {\bibfnamefont {Y.}~\bibnamefont {Wang}},\ }\href {\doibase
  10.1088/1475-7516/2021/03/009} {\bibfield  {journal} {\bibinfo  {journal}
  {JCAP}\ }\textbf {\bibinfo {volume} {03}},\ \bibinfo {pages} {009} (\bibinfo
  {year} {2021})},\ \Eprint {http://arxiv.org/abs/2007.10343} {arXiv:2007.10343
  [hep-ph]} \BibitemShut {NoStop}%
\bibitem [{\citenamefont {Azatov}\ and\ \citenamefont
  {Vanvlasselaer}(2021)}]{Azatov:2020ufh}%
  \BibitemOpen
  \bibfield  {author} {\bibinfo {author} {\bibfnamefont {A.}~\bibnamefont
  {Azatov}}\ and\ \bibinfo {author} {\bibfnamefont {M.}~\bibnamefont
  {Vanvlasselaer}},\ }\href {\doibase 10.1088/1475-7516/2021/01/058} {\bibfield
   {journal} {\bibinfo  {journal} {JCAP}\ }\textbf {\bibinfo {volume} {01}},\
  \bibinfo {pages} {058} (\bibinfo {year} {2021})},\ \Eprint
  {http://arxiv.org/abs/2010.02590} {arXiv:2010.02590 [hep-ph]} \BibitemShut
  {NoStop}%
\bibitem [{\citenamefont {Garcia~Garcia}\ \emph {et~al.}(2023)\citenamefont
  {Garcia~Garcia}, \citenamefont {Koszegi},\ and\ \citenamefont
  {Petrossian-Byrne}}]{GarciaGarcia:2022yqb}%
  \BibitemOpen
  \bibfield  {author} {\bibinfo {author} {\bibfnamefont {I.}~\bibnamefont
  {Garcia~Garcia}}, \bibinfo {author} {\bibfnamefont {G.}~\bibnamefont
  {Koszegi}}, \ and\ \bibinfo {author} {\bibfnamefont {R.}~\bibnamefont
  {Petrossian-Byrne}},\ }\href {\doibase 10.1007/JHEP09(2023)013} {\bibfield
  {journal} {\bibinfo  {journal} {JHEP}\ }\textbf {\bibinfo {volume} {09}},\
  \bibinfo {pages} {013} (\bibinfo {year} {2023})},\ \Eprint
  {http://arxiv.org/abs/2212.10572} {arXiv:2212.10572 [hep-ph]} \BibitemShut
  {NoStop}%
\bibitem [{\citenamefont {Azatov}\ \emph {et~al.}(2023)\citenamefont {Azatov},
  \citenamefont {Barni}, \citenamefont {Petrossian-Byrne},\ and\ \citenamefont
  {Vanvlasselaer}}]{Azatov:2023xem}%
  \BibitemOpen
  \bibfield  {author} {\bibinfo {author} {\bibfnamefont {A.}~\bibnamefont
  {Azatov}}, \bibinfo {author} {\bibfnamefont {G.}~\bibnamefont {Barni}},
  \bibinfo {author} {\bibfnamefont {R.}~\bibnamefont {Petrossian-Byrne}}, \
  and\ \bibinfo {author} {\bibfnamefont {M.}~\bibnamefont {Vanvlasselaer}},\
  }\href@noop {} {\  (\bibinfo {year} {2023})},\ \Eprint
  {http://arxiv.org/abs/2310.06972} {arXiv:2310.06972 [hep-ph]} \BibitemShut
  {NoStop}%
\bibitem [{\citenamefont {Gouttenoire}\ \emph {et~al.}(2023)\citenamefont
  {Gouttenoire}, \citenamefont {Kuflik},\ and\ \citenamefont
  {Liu}}]{Gouttenoire:2023roe}%
  \BibitemOpen
  \bibfield  {author} {\bibinfo {author} {\bibfnamefont {Y.}~\bibnamefont
  {Gouttenoire}}, \bibinfo {author} {\bibfnamefont {E.}~\bibnamefont {Kuflik}},
  \ and\ \bibinfo {author} {\bibfnamefont {D.}~\bibnamefont {Liu}},\
  }\href@noop {} {\bibfield  {journal} {\bibinfo  {journal} {arXiv preprint}\ }
  (\bibinfo {year} {2023})},\ \Eprint {http://arxiv.org/abs/2311.00029}
  {arXiv:2311.00029 [hep-ph]} \BibitemShut {NoStop}%
\bibitem [{\citenamefont {Borsanyi}\ \emph {et~al.}(2023)\citenamefont
  {Borsanyi}, \citenamefont {Fodor}, \citenamefont {Godzieba}, \citenamefont
  {Kara}, \citenamefont {Parotto}, \citenamefont {Sexty},\ and\ \citenamefont
  {Vig}}]{Borsanyi:2022fub}%
  \BibitemOpen
  \bibfield  {author} {\bibinfo {author} {\bibfnamefont {S.}~\bibnamefont
  {Borsanyi}}, \bibinfo {author} {\bibfnamefont {Z.}~\bibnamefont {Fodor}},
  \bibinfo {author} {\bibfnamefont {D.~A.}\ \bibnamefont {Godzieba}}, \bibinfo
  {author} {\bibfnamefont {R.}~\bibnamefont {Kara}}, \bibinfo {author}
  {\bibfnamefont {P.}~\bibnamefont {Parotto}}, \bibinfo {author} {\bibfnamefont
  {D.}~\bibnamefont {Sexty}}, \ and\ \bibinfo {author} {\bibfnamefont
  {R.}~\bibnamefont {Vig}},\ }\href {\doibase 10.1103/PhysRevD.107.054514}
  {\bibfield  {journal} {\bibinfo  {journal} {Phys. Rev. D}\ }\textbf {\bibinfo
  {volume} {107}},\ \bibinfo {pages} {054514} (\bibinfo {year} {2023})},\
  \Eprint {http://arxiv.org/abs/2212.08684} {arXiv:2212.08684 [hep-lat]}
  \BibitemShut {NoStop}%
\bibitem [{\citenamefont {Sarkar}\ \emph {et~al.}(2015)\citenamefont {Sarkar},
  \citenamefont {Das},\ and\ \citenamefont {Sethi}}]{Sarkar:2014bca}%
  \BibitemOpen
  \bibfield  {author} {\bibinfo {author} {\bibfnamefont {A.}~\bibnamefont
  {Sarkar}}, \bibinfo {author} {\bibfnamefont {S.}~\bibnamefont {Das}}, \ and\
  \bibinfo {author} {\bibfnamefont {S.~K.}\ \bibnamefont {Sethi}},\ }\href
  {\doibase 10.1088/1475-7516/2015/03/004} {\bibfield  {journal} {\bibinfo
  {journal} {JCAP}\ }\textbf {\bibinfo {volume} {03}},\ \bibinfo {pages} {004}
  (\bibinfo {year} {2015})},\ \Eprint {http://arxiv.org/abs/1410.7129}
  {arXiv:1410.7129 [astro-ph.CO]} \BibitemShut {NoStop}%
\bibitem [{\citenamefont {Husdal}(2016)}]{Husdal:2016haj}%
  \BibitemOpen
  \bibfield  {author} {\bibinfo {author} {\bibfnamefont {L.}~\bibnamefont
  {Husdal}},\ }\href {\doibase 10.3390/galaxies4040078} {\bibfield  {journal}
  {\bibinfo  {journal} {Galaxies}\ }\textbf {\bibinfo {volume} {4}},\ \bibinfo
  {pages} {78} (\bibinfo {year} {2016})},\ \Eprint
  {http://arxiv.org/abs/1609.04979} {arXiv:1609.04979 [astro-ph.CO]}
  \BibitemShut {NoStop}%
\bibitem [{\citenamefont {Hardy}(2017)}]{Hardy:2016mns}%
  \BibitemOpen
  \bibfield  {author} {\bibinfo {author} {\bibfnamefont {E.}~\bibnamefont
  {Hardy}},\ }\href {\doibase 10.1007/JHEP02(2017)046} {\bibfield  {journal}
  {\bibinfo  {journal} {JHEP}\ }\textbf {\bibinfo {volume} {02}},\ \bibinfo
  {pages} {046} (\bibinfo {year} {2017})},\ \Eprint
  {http://arxiv.org/abs/1609.00208} {arXiv:1609.00208 [hep-ph]} \BibitemShut
  {NoStop}%
\bibitem [{\citenamefont {Bogolyubsky}\ and\ \citenamefont
  {Makhankov}(1976)}]{Bogolyubsky:1976nx}%
  \BibitemOpen
  \bibfield  {author} {\bibinfo {author} {\bibfnamefont {I.~L.}\ \bibnamefont
  {Bogolyubsky}}\ and\ \bibinfo {author} {\bibfnamefont {V.~G.}\ \bibnamefont
  {Makhankov}},\ }\href@noop {} {\bibfield  {journal} {\bibinfo  {journal}
  {JETP Lett.}\ }\textbf {\bibinfo {volume} {24}},\ \bibinfo {pages} {12}
  (\bibinfo {year} {1976})}\BibitemShut {NoStop}%
\bibitem [{\citenamefont {Segur}\ and\ \citenamefont
  {Kruskal}(1987)}]{Segur:1987mg}%
  \BibitemOpen
  \bibfield  {author} {\bibinfo {author} {\bibfnamefont {H.}~\bibnamefont
  {Segur}}\ and\ \bibinfo {author} {\bibfnamefont {M.~D.}\ \bibnamefont
  {Kruskal}},\ }\href {\doibase 10.1103/PhysRevLett.58.747} {\bibfield
  {journal} {\bibinfo  {journal} {Phys. Rev. Lett.}\ }\textbf {\bibinfo
  {volume} {58}},\ \bibinfo {pages} {747} (\bibinfo {year} {1987})}\BibitemShut
  {NoStop}%
\bibitem [{\citenamefont {Gleiser}(1994)}]{Gleiser:1993pt}%
  \BibitemOpen
  \bibfield  {author} {\bibinfo {author} {\bibfnamefont {M.}~\bibnamefont
  {Gleiser}},\ }\href {\doibase 10.1103/PhysRevD.49.2978} {\bibfield  {journal}
  {\bibinfo  {journal} {Phys. Rev. D}\ }\textbf {\bibinfo {volume} {49}},\
  \bibinfo {pages} {2978} (\bibinfo {year} {1994})},\ \Eprint
  {http://arxiv.org/abs/hep-ph/9308279} {arXiv:hep-ph/9308279} \BibitemShut
  {NoStop}%
\bibitem [{\citenamefont {Copeland}\ \emph {et~al.}(1995)\citenamefont
  {Copeland}, \citenamefont {Gleiser},\ and\ \citenamefont
  {Muller}}]{Copeland:1995fq}%
  \BibitemOpen
  \bibfield  {author} {\bibinfo {author} {\bibfnamefont {E.~J.}\ \bibnamefont
  {Copeland}}, \bibinfo {author} {\bibfnamefont {M.}~\bibnamefont {Gleiser}}, \
  and\ \bibinfo {author} {\bibfnamefont {H.~R.}\ \bibnamefont {Muller}},\
  }\href {\doibase 10.1103/PhysRevD.52.1920} {\bibfield  {journal} {\bibinfo
  {journal} {Phys. Rev. D}\ }\textbf {\bibinfo {volume} {52}},\ \bibinfo
  {pages} {1920} (\bibinfo {year} {1995})},\ \Eprint
  {http://arxiv.org/abs/hep-ph/9503217} {arXiv:hep-ph/9503217} \BibitemShut
  {NoStop}%
\bibitem [{\citenamefont {Gleiser}\ and\ \citenamefont
  {Sornborger}(2000)}]{Gleiser:1999tj}%
  \BibitemOpen
  \bibfield  {author} {\bibinfo {author} {\bibfnamefont {M.}~\bibnamefont
  {Gleiser}}\ and\ \bibinfo {author} {\bibfnamefont {A.}~\bibnamefont
  {Sornborger}},\ }\href {\doibase 10.1103/PhysRevE.62.1368} {\bibfield
  {journal} {\bibinfo  {journal} {Phys. Rev. E}\ }\textbf {\bibinfo {volume}
  {62}},\ \bibinfo {pages} {1368} (\bibinfo {year} {2000})},\ \Eprint
  {http://arxiv.org/abs/patt-sol/9909002} {arXiv:patt-sol/9909002} \BibitemShut
  {NoStop}%
\bibitem [{\citenamefont {Honda}\ and\ \citenamefont
  {Choptuik}(2002)}]{Honda:2001xg}%
  \BibitemOpen
  \bibfield  {author} {\bibinfo {author} {\bibfnamefont {E.~P.}\ \bibnamefont
  {Honda}}\ and\ \bibinfo {author} {\bibfnamefont {M.~W.}\ \bibnamefont
  {Choptuik}},\ }\href {\doibase 10.1103/PhysRevD.65.084037} {\bibfield
  {journal} {\bibinfo  {journal} {Phys. Rev. D}\ }\textbf {\bibinfo {volume}
  {65}},\ \bibinfo {pages} {084037} (\bibinfo {year} {2002})},\ \Eprint
  {http://arxiv.org/abs/hep-ph/0110065} {arXiv:hep-ph/0110065} \BibitemShut
  {NoStop}%
\bibitem [{\citenamefont {Kasuya}\ \emph {et~al.}(2003)\citenamefont {Kasuya},
  \citenamefont {Kawasaki},\ and\ \citenamefont {Takahashi}}]{Kasuya:2002zs}%
  \BibitemOpen
  \bibfield  {author} {\bibinfo {author} {\bibfnamefont {S.}~\bibnamefont
  {Kasuya}}, \bibinfo {author} {\bibfnamefont {M.}~\bibnamefont {Kawasaki}}, \
  and\ \bibinfo {author} {\bibfnamefont {F.}~\bibnamefont {Takahashi}},\ }\href
  {\doibase 10.1016/S0370-2693(03)00344-7} {\bibfield  {journal} {\bibinfo
  {journal} {Phys. Lett. B}\ }\textbf {\bibinfo {volume} {559}},\ \bibinfo
  {pages} {99} (\bibinfo {year} {2003})},\ \Eprint
  {http://arxiv.org/abs/hep-ph/0209358} {arXiv:hep-ph/0209358} \BibitemShut
  {NoStop}%
\bibitem [{\citenamefont {Kitajima}\ and\ \citenamefont
  {Takahashi}(2023)}]{Kitajima:2023pby}%
  \BibitemOpen
  \bibfield  {author} {\bibinfo {author} {\bibfnamefont {N.}~\bibnamefont
  {Kitajima}}\ and\ \bibinfo {author} {\bibfnamefont {F.}~\bibnamefont
  {Takahashi}},\ }\href {\doibase 10.1103/PhysRevD.107.123518} {\bibfield
  {journal} {\bibinfo  {journal} {Phys. Rev. D}\ }\textbf {\bibinfo {volume}
  {107}},\ \bibinfo {pages} {123518} (\bibinfo {year} {2023})},\ \Eprint
  {http://arxiv.org/abs/2303.05492} {arXiv:2303.05492 [hep-ph]} \BibitemShut
  {NoStop}%
\bibitem [{\citenamefont {Agrawal}\ \emph {et~al.}(2020)\citenamefont
  {Agrawal}, \citenamefont {Kitajima}, \citenamefont {Reece}, \citenamefont
  {Sekiguchi},\ and\ \citenamefont {Takahashi}}]{Agrawal:2018vin}%
  \BibitemOpen
  \bibfield  {author} {\bibinfo {author} {\bibfnamefont {P.}~\bibnamefont
  {Agrawal}}, \bibinfo {author} {\bibfnamefont {N.}~\bibnamefont {Kitajima}},
  \bibinfo {author} {\bibfnamefont {M.}~\bibnamefont {Reece}}, \bibinfo
  {author} {\bibfnamefont {T.}~\bibnamefont {Sekiguchi}}, \ and\ \bibinfo
  {author} {\bibfnamefont {F.}~\bibnamefont {Takahashi}},\ }\href {\doibase
  10.1016/j.physletb.2019.135136} {\bibfield  {journal} {\bibinfo  {journal}
  {Phys. Lett. B}\ }\textbf {\bibinfo {volume} {801}},\ \bibinfo {pages}
  {135136} (\bibinfo {year} {2020})},\ \Eprint
  {http://arxiv.org/abs/1810.07188} {arXiv:1810.07188 [hep-ph]} \BibitemShut
  {NoStop}%
\bibitem [{\citenamefont {Co}\ \emph {et~al.}(2019)\citenamefont {Co},
  \citenamefont {Pierce}, \citenamefont {Zhang},\ and\ \citenamefont
  {Zhao}}]{Co:2018lka}%
  \BibitemOpen
  \bibfield  {author} {\bibinfo {author} {\bibfnamefont {R.~T.}\ \bibnamefont
  {Co}}, \bibinfo {author} {\bibfnamefont {A.}~\bibnamefont {Pierce}}, \bibinfo
  {author} {\bibfnamefont {Z.}~\bibnamefont {Zhang}}, \ and\ \bibinfo {author}
  {\bibfnamefont {Y.}~\bibnamefont {Zhao}},\ }\href {\doibase
  10.1103/PhysRevD.99.075002} {\bibfield  {journal} {\bibinfo  {journal} {Phys.
  Rev. D}\ }\textbf {\bibinfo {volume} {99}},\ \bibinfo {pages} {075002}
  (\bibinfo {year} {2019})},\ \Eprint {http://arxiv.org/abs/1810.07196}
  {arXiv:1810.07196 [hep-ph]} \BibitemShut {NoStop}%
\bibitem [{\citenamefont {Bastero-Gil}\ \emph {et~al.}(2019)\citenamefont
  {Bastero-Gil}, \citenamefont {Santiago}, \citenamefont {Ubaldi},\ and\
  \citenamefont {Vega-Morales}}]{Bastero-Gil:2018uel}%
  \BibitemOpen
  \bibfield  {author} {\bibinfo {author} {\bibfnamefont {M.}~\bibnamefont
  {Bastero-Gil}}, \bibinfo {author} {\bibfnamefont {J.}~\bibnamefont
  {Santiago}}, \bibinfo {author} {\bibfnamefont {L.}~\bibnamefont {Ubaldi}}, \
  and\ \bibinfo {author} {\bibfnamefont {R.}~\bibnamefont {Vega-Morales}},\
  }\href {\doibase 10.1088/1475-7516/2019/04/015} {\bibfield  {journal}
  {\bibinfo  {journal} {JCAP}\ }\textbf {\bibinfo {volume} {04}},\ \bibinfo
  {pages} {015} (\bibinfo {year} {2019})},\ \Eprint
  {http://arxiv.org/abs/1810.07208} {arXiv:1810.07208 [hep-ph]} \BibitemShut
  {NoStop}%
\bibitem [{\citenamefont {Minami}\ and\ \citenamefont
  {Komatsu}(2020)}]{Minami:2020odp}%
  \BibitemOpen
  \bibfield  {author} {\bibinfo {author} {\bibfnamefont {Y.}~\bibnamefont
  {Minami}}\ and\ \bibinfo {author} {\bibfnamefont {E.}~\bibnamefont
  {Komatsu}},\ }\href {\doibase 10.1103/PhysRevLett.125.221301} {\bibfield
  {journal} {\bibinfo  {journal} {Phys. Rev. Lett.}\ }\textbf {\bibinfo
  {volume} {125}},\ \bibinfo {pages} {221301} (\bibinfo {year} {2020})},\
  \Eprint {http://arxiv.org/abs/2011.11254} {arXiv:2011.11254 [astro-ph.CO]}
  \BibitemShut {NoStop}%
\bibitem [{\citenamefont {Diego-Palazuelos}\ \emph {et~al.}(2022)\citenamefont
  {Diego-Palazuelos} \emph {et~al.}}]{Diego-Palazuelos:2022dsq}%
  \BibitemOpen
  \bibfield  {author} {\bibinfo {author} {\bibfnamefont {P.}~\bibnamefont
  {Diego-Palazuelos}} \emph {et~al.},\ }\href {\doibase
  10.1103/PhysRevLett.128.091302} {\bibfield  {journal} {\bibinfo  {journal}
  {Phys. Rev. Lett.}\ }\textbf {\bibinfo {volume} {128}},\ \bibinfo {pages}
  {091302} (\bibinfo {year} {2022})},\ \Eprint
  {http://arxiv.org/abs/2201.07682} {arXiv:2201.07682 [astro-ph.CO]}
  \BibitemShut {NoStop}%
\bibitem [{\citenamefont {Eskilt}(2022)}]{Eskilt:2022wav}%
  \BibitemOpen
  \bibfield  {author} {\bibinfo {author} {\bibfnamefont {J.~R.}\ \bibnamefont
  {Eskilt}},\ }\href {\doibase 10.1051/0004-6361/202243269} {\bibfield
  {journal} {\bibinfo  {journal} {Astron. Astrophys.}\ }\textbf {\bibinfo
  {volume} {662}},\ \bibinfo {pages} {A10} (\bibinfo {year} {2022})},\ \Eprint
  {http://arxiv.org/abs/2201.13347} {arXiv:2201.13347 [astro-ph.CO]}
  \BibitemShut {NoStop}%
\bibitem [{\citenamefont {Eskilt}\ and\ \citenamefont
  {Komatsu}(2022)}]{Eskilt:2022cff}%
  \BibitemOpen
  \bibfield  {author} {\bibinfo {author} {\bibfnamefont {J.~R.}\ \bibnamefont
  {Eskilt}}\ and\ \bibinfo {author} {\bibfnamefont {E.}~\bibnamefont
  {Komatsu}},\ }\href {\doibase 10.1103/PhysRevD.106.063503} {\bibfield
  {journal} {\bibinfo  {journal} {Phys. Rev. D}\ }\textbf {\bibinfo {volume}
  {106}},\ \bibinfo {pages} {063503} (\bibinfo {year} {2022})},\ \Eprint
  {http://arxiv.org/abs/2205.13962} {arXiv:2205.13962 [astro-ph.CO]}
  \BibitemShut {NoStop}%
\bibitem [{\citenamefont {Witten}(1980)}]{Witten:1980sp}%
  \BibitemOpen
  \bibfield  {author} {\bibinfo {author} {\bibfnamefont {E.}~\bibnamefont
  {Witten}},\ }\href {\doibase 10.1016/0003-4916(80)90325-5} {\bibfield
  {journal} {\bibinfo  {journal} {Annals Phys.}\ }\textbf {\bibinfo {volume}
  {128}},\ \bibinfo {pages} {363} (\bibinfo {year} {1980})}\BibitemShut
  {NoStop}%
\bibitem [{\citenamefont {Azatov}\ \emph
  {et~al.}(2021{\natexlab{b}})\citenamefont {Azatov}, \citenamefont
  {Vanvlasselaer},\ and\ \citenamefont {Yin}}]{Azatov:2021irb}%
  \BibitemOpen
  \bibfield  {author} {\bibinfo {author} {\bibfnamefont {A.}~\bibnamefont
  {Azatov}}, \bibinfo {author} {\bibfnamefont {M.}~\bibnamefont
  {Vanvlasselaer}}, \ and\ \bibinfo {author} {\bibfnamefont {W.}~\bibnamefont
  {Yin}},\ }\href {\doibase 10.1007/JHEP10(2021)043} {\bibfield  {journal}
  {\bibinfo  {journal} {JHEP}\ }\textbf {\bibinfo {volume} {10}},\ \bibinfo
  {pages} {043} (\bibinfo {year} {2021}{\natexlab{b}})},\ \Eprint
  {http://arxiv.org/abs/2106.14913} {arXiv:2106.14913 [hep-ph]} \BibitemShut
  {NoStop}%
\end{thebibliography}%

\end{document}